\documentclass[twocolumn,amsmath,amssymb]{revtex4}
\usepackage{amsfonts}
\usepackage[usenames]{color}

\newcommand{\be}{\begin{equation}}
\newcommand{\ee}{\end{equation}}
\newcommand{\bea}{\begin{eqnarray}}
\newcommand{\eea}{\end{eqnarray}}

\begin{document}
\title{Cosmological perturbations of a relativistic MOND theory}
\author{Jai-chan Hwang${}^{1}$ and Hyerim Noh${}^{2}$}
\address{${}^{1}$Particle Theory  and Cosmology Group,
         Center for Theoretical Physics of the Universe,
         Institute for Basic Science (IBS), Daejeon, 34051, Republic of Korea
         \\
         ${}^{2}$Theoretical Astrophysics Group, Korea Astronomy and Space Science Institute, Daejeon, Republic of Korea
         }


\begin{abstract}

A relativistic MOND theory, promising in reproducing cosmology as well as the MOND phenomenology in the low acceleration regime, was recently proposed. We present the post-Newtonian (PN) approximation and relativistic perturbation equations of this theory in cosmological context. The PN equations are presented to 1PN order and perturbation equations are presented in fully-nonlinear and exact forms. The gauge issues are clarified. The 1PN equations and linear perturbation equations are presented without imposing temporal gauge conditions. We show that to 0PN order baryon perturbation grows faster in the MOND regime. The MOND field can be interpreted as a fluid with specific equation of state without anisotropic stress, and the Jeans criterion is derived for the MOND field.

\end{abstract}

\maketitle

\tableofcontents

%
%
%
\section{Introduction}
                                        \label{sec:introduction}

Milgrom, in 1983, proposed MOND (Modified Newtonian Dynamics) to explain the gravity anomaly in galaxy rotations \cite{MOND-1983a}. Instead of a popular way of saving conventional gravity (both Newton's and Einstein's) by introducing unseen dark matter (DM), he has modified in a non-relativistic context either gravity or inertia for the anomaly. Despite the lack of agreed relativistic theory, thus unable to address the cosmological demands other than the galactic rotation and related phenomena on similar scales, the non-relativistic proposal has survived till today as an alternative to the dark matter hypothesis. Recently, it gained independent observational support and challenge in stellar and Solar system scale phenomena; see \cite{Chae-2023, Hernandez-2023, Chae-2024a, Chae-2024b, Hernandez-2024, Kroupa-2024} for support from the long period binary stars observed by GAIA and from tidal tail studies of open clusters, and \cite{Vokrouhlicky-2024, Desmond-2024} for challenges to specific MOND models in explaining the gravitational field and the long period dynamics in the solar system.

Such independent tests of the proposal in stellar and solar system scales are possible because Milgrom has proposed modification of gravity or inertia in the low acceleration regime by $f(|\vec{a}|/a_M) \vec{a} = \vec{g}_N$ where $\vec{g}_N$ is the Newtonian gravitational acceleration, with $f(x) = 1$ for $x \gg 1$, thus recovering Newton's gravity, and $f(x) = x$ for $x \ll 1$ in the low acceleration MOND regime. In the MOND regime, we have $|\vec{a}| = \sqrt{a_M g_N}$ with $g_N = |\vec{g}_N|$. For an isolated spherically symmetric system with $g_N = GM/r^2$, using $|\vec{a}| = v^2/r$, we have $v = (G M a_M)^{1/4} \propto M^{1/4}$ thus naturally recovering the observationally known galaxies' flat rotation curves and Tully-Fisher/Faber-Jackson relation \cite{Sanders-McGaugh-2002}.

Observations have pin down the Milgrom's universal constant of nature $a_M \simeq 1.2 \times 10^{-8} cm/sec^2$ \cite{McGaugh-2004}. As originally noticed by Milgrom \cite{MOND-1983a} the transition acceleration $a_M$ is intriguingly close to the acceleration scale of the expanding Universe $c H_0/(2 \pi) \sim 1.1 \times 10^{-8} cm/sec^2$ or $c^2 \sqrt{\Lambda} = \sqrt{3 \Omega_{\Lambda 0}} c H_0 \sim 1.4 c H_0$ using the current value of Lema\^itre-Hubble parameter $H_0 \sim 70 (km/sec)/Mpc$ and $\Omega_{\Lambda 0} \sim 0.69$; we have $c H_0 = L_H/t_H^2$ with $L_H \equiv c/H_0$ and $t_H \equiv 1/H_0$. The interpolating function $f$ is not determined yet.

Bekenstein and Milgrom, in 1984, proposed a non-relativistic Lagrangian formulation of MOND, often called as AQUAL, leading to a modified Poisson's equation $\nabla \cdot (f \nabla \Phi) = 4 \pi G \varrho$ with $\vec{a} = \nabla \Phi$ \cite{Bekenstein-Milgrom-1984}. At large distance from an isolated body with high symmetry, Milgrom's original proposal is recovered \cite{Bekenstein-Milgrom-1984}. In the MOND based on the modified Poisson's equation, the internal gravitational dynamics of a system is affected by an external gravitational field, often called the external-field effect, thus violating the strong equivalence principle \cite{Bekenstein-Milgrom-1984}. A variety of alternatives to AQUAL theory producing the MOND phenomenology is suggested \cite{Milgrom-2010-QUMOND, Milgrom-2023-TriMOND, Milgrom-2023-GQUMOND}, and diverse relativistic extensions are also proposed \cite{Bekenstein-Milgrom-1984, Famaey-2012}.

Blanchet and Marsat \cite{Blanchet-Marsat-2011} introduced a vector field {\it in} Einstein's gravity to recover the MOND phenomenology in weak gravity low acceleration regime. They used a function of the acceleration of a four-vector $U_a$ associated with a scalar function $\tau$ as $U_a \propto \nabla_a \tau$. The theory is motivated by a quantum gravity proposal with local Lorentz invariance violation \cite{Horava-2009, Blas-2010} and Einstein aether theories \cite{Jacobson-Mattingly-2001, Zlosnik-2007, Jacobson-2010, Sanders-2011}.

The MOND proposal is successful in explaining the phenomenology in the galactic scale without demanding DM. However, successful cosmology also demands the presence of a sort of DM (a fluid with dust-like equation of state) in explaining phenomena in the large-scale structure, the cosmic background radiation (CMB), and the background matter content. Blanchet and Skordis \cite{Blanchet-Skordis-2024} recently proposed one more additional function in the Blanchet and Marsat's model, now using a function of the normalization amplitude of {\it the same} four-vector $U_a$ to accommodate such demands in cosmology.

The theory introduces two free functions of the $\tau$-field (MOND field): a function of the acceleration ${\cal J}(A)$ of a unit four-vector $U_a$ for the MOND phenomenology in low acceleration regime and the other a function of the normalization amplitude ${\cal K}(Q)$ of the same four-vector for successful cosmology, see Eq.\ (\ref{L}). Forms of the two functions are still free to accommodate the phenomena in cosmology and MOND \cite{Blanchet-Skordis-2024}. Future studies may guide or design the functional forms of ${\cal J}(A)$ and ${\cal K}(Q)$ using theories and observations. For such studies the cosmological background and perturbation equations are needed.

Here, we present equations for the post-Newtonian (PN) approximation and relativistic perturbation theory in cosmological context. The PN approximation is presented to 1PN order, and the perturbation theory is presented in fully-nonlinear and exact form. We address the gauge issues in both formulations and the linear perturbation equations are presented without imposing temporal gauge conditions. These subjects are already studied in \cite{Blanchet-Skordis-2024} and here we elaborate some aspects. For both formulations, we compare the MOND contribution with an ordinary fluid. To the linear order in perturbation, the MOND field does not contribute to the vector- and tensor-type perturbations, and it can be interpreted as a fluid with a specific equation of state (depending on the functional forms of ${\cal J}$ and ${\cal K}$) and is not associated with the anisotropic stress.

Section \ref{sec:covariant} summarizes the Lagrangian formulation of the theory with Einstein's equation and the equation of motion in covariant forms. Sections \ref{sec:BG}-\ref{sec:PT} derive cosmological background equations, the cosmological PN equations to 1PN order, and the cosmological perturbation equations valid in fully-nonlinear and exact form, respectively. Section \ref{sec:discussion} is a discussion.

In the post-Newtonian approximation we use $x^0 = ct$ for derivation, whereas in the background and perturbation theory we use the conformal time $x^0 = \eta$ with $c dt = a d\eta$; $a$ is the cosmic scale factor.

%
%
%
\section{Relativistic MOND}
                                        \label{sec:covariant}

Blanchet, Marsat and Skordis \cite{Blanchet-Marsat-2011, Blanchet-Skordis-2024} introduced the following Lagrangian as a relativistic extension of the MOND proposal using a $\tau$-field
\bea
   {\cal L} = \sqrt{-g} \Big\{ {c^4 \over 16 \pi G}
       \big[ R - 2 {\cal J} (A) + 2 {\cal K} (Q) \big]
       + L_{\rm m} \Big\},
   \label{L}
\eea
where
\bea
   & & \hskip -.8cm
       U_a \equiv - {c \over Q} \nabla_a \tau, \quad
       Q \equiv c \sqrt{- (\nabla^c \tau) \nabla_c \tau},
   \nonumber \\
   & & \hskip -.8cm
       A_a \equiv c^2 U_{a;b} U^b = - c^2 q^b_a \nabla_b \ln{Q}, \quad
       A \equiv {1 \over c^4} A^c A_c,
   \label{Q}
\eea
and $L_m$ is the ordinary matter (including baryon and photon) part Lagrangian. Although we call the fundamental field simply a $\tau$-field, what is introduced is a four-vector field $U_a$ defined using the $\tau$-field; $U_a$ is a normalized timelike four-vector with $U^c U_c \equiv - 1$; $q_{ab} \equiv g_{ab} + U_a U_b$ is a spatial projection tensor orthogonal to $U_a$; $A_a$ is an acceleration of the four-vector. For $\tau = \tau(t)$, we have $U_a = n_a$ where $n_a$ is the normal four-vector with $n_i \equiv 0$.

The ${\cal J}$-term is used to realize the MOND phenomenology in low acceleration regime by Blanchet and Marsat \cite{Blanchet-Marsat-2011}. The ${\cal K}$-term is added in order to accommodate the cosmological demands of the dark matter in CMB and the large-scale structure using the same $\tau$-field by Blanchet and Skordis \cite{Blanchet-Skordis-2024}. We may call it Blanchet-Marsat-Skordis (BMS) theory. The cosmological constant can be absorbed to ${\cal J}$ or ${\cal K}$ as ${\cal J} = \Lambda$ or ${\cal K} = - \Lambda$; here we consider the cosmological constant separately in $L_m$.

Variation with respect to $g_{ab}$ gives
\bea
   G_{ab} = {8 \pi G \over c^4} T_{ab},
\eea
where the ${\cal J}$ and ${\cal K}$-parts are
\bea
   & & \hskip -.5cm
       T^{\cal J}_{ab}
       = {c^4 \over 8 \pi G} \Big[ - {\cal J} g_{ab}
       + {2 \over c^4} {\cal J}_{,A} A_a A_b
       - {2 \over c^2} ( {\cal J}_{,A} A^c )_{;c}
       U_a U_b \Big],
   \nonumber \\
   & & \hskip -.5cm
       T^{\cal K}_{ab}
       = {c^4 \over 8 \pi G} ( {\cal K} g_{ab}
       + Q {\cal K}_{,Q} U_a U_b ).
   \label{Tab}
\eea
We will call the sum of these two parts the $\tau$-field (or $\tau$-fluid) energy-momentum tensor. We have $T_{ab} = T_{ab}^m + T_{ab}^\tau$ with $T_{ab}^m$ the matter part. The covariant fluid quantities measured by an observer with a general four-vector $u_a$ are introduced as \cite{Ellis-1971}
\bea
   & & \hskip -1cm
       T_{ab} = \mu u_a u_b
       + p \left( g_{ab} + u_a u_b \right)
       + q_a u_b + q_b u_a + \pi_{ab},
   \label{Tab-fluid} \\
   & & \hskip -1cm
       \mu = T_{ab} u^a u^b, \quad
       p = {1 \over 3} T_{ab} h^{ab}, \quad
       q_a = - T_{cd} u^c h^d_a,
   \nonumber \\
   & & \hskip -1cm
       \pi_{ab} = T_{cd} h^c_a h^d_b - p h_{ab},
   \label{fluid-quantities}
\eea
where $h_{ab} \equiv g_{ab} + u_a u_b$ is a spatial projection tensor orthogonal to $u_a$.

Variation with respect to $\tau$ gives the equation of motion
\bea
   S^a_{\;\; ;a} = 0,
   \label{EOM}
\eea
with
\bea
   S^a \equiv {2 \over Q} {\cal J}_{,A} A^a
       U^b \nabla_b \ln{Q}
       + \Big[ {2 \over Q} ( {\cal J}_{,A} A^b )_{;b}
       - {\cal K}_{,Q} c^2 \Big] U^a.
   \label{Sa}
\eea
Using the equation of motion, we can show $(T^{ab}_{\cal J} + T^{ab}_{\cal K})_{;b} = 0$.

The MOND phenomenology will constrain the form of ${\cal J}$. We will often consider two models for ${\cal K}$ \cite{Blanchet-Skordis-2024}
\bea
   & & \hskip -.8cm
       {\rm Model \; I}: \hskip .13cm
       {\cal K} = {2 \nu^2 \over n + 1}
       {\cal K}_{n+1} ( Q - 1 )^{n+1},
   \label{model-I} \\
   & & \hskip -.8cm
       {\rm Model \; II}:
       {\cal K} = \sum_{n = 1}^\infty {2 \nu^2 \over n + 1}
       {\cal K}_{n+1} ( Q - 1 )^{n+1},
   \label{model-II}
\eea
with ${\cal K}_2 \equiv 1$.

%
%
%
\section{Cosmological Background}
                                        \label{sec:BG}

We consider a flat Friedmann cosmology with the Robertson-Walker metric
\bea
   ds^2 = - a^2 d x^0 d x^0 + a^2 \delta_{ij} dx^i dx^j,
\eea
where $x^0 = \eta$ is the conformal time. Using $\tau = \tau(t)$, we have
\bea
   & & U_0 = - a, \quad
       U_i = 0, \quad
       U^0 = {1 \over a}, \quad
       U^i = 0;
   \nonumber \\
   & & Q = \dot \tau, \quad A_a = 0 = A,
\eea
where an overdot indicates a derivative with respect to the cosmic time $t$. Equation (\ref{Tab}), with $T_{00} = a^2 \mu$ and $T_{ij} = a^2 p \delta_{ij}$, gives $\tau$-fluid quantities
\bea
   \mu_\tau = {c^4 \over 8 \pi G} ( {\cal J} - {\cal K}
       + Q {\cal K}_{,Q} ), \quad
       p_\tau = {c^4 \over 8 \pi G} ( - {\cal J} + {\cal K} ).
   \label{fluid-BG}
\eea
As we have $A = 0$, only ${\cal J} = {\rm constant}$ is allowed which is absorbed to the cosmological constant; thus, we set ${\cal J} = 0$.

Einstein's equation gives
\bea
   & & \hskip -1.cm
       H^2 = {8 \pi G \over 3 c^2} \mu_m
       - {c^2 \over 3} ( {\cal K} - Q {\cal K}_{,Q} )
       + {\Lambda c^2 \over 3},
   \label{BG-eq1} \\
   & & \hskip -1.cm
       {\ddot a \over a} = - {4 \pi G \over 3 c^2}
       ( \mu_m + 3 p_m )
       - {c^2 \over 3} \Big( {\cal K}
       + {1 \over 2} Q {\cal K}_{,Q} \Big)
       + {\Lambda c^2 \over 3},
   \label{BG-eq2}
\eea
where $\mu_m$ and $p_m$ are the fluid part of energy density and pressure, respectively; these are sum of radiation, baryon, etc., with respective conservation equations $\dot \mu_m + 3 H (\mu_m + p_m) = 0$; $H \equiv \dot a/a$.

The equation of motion gives $( a^3 {\cal K}_{,Q} )^\prime = 0$, thus \cite{Blanchet-Skordis-2024}
\bea
   {\cal K}_{,Q} = {I_0 \over a^3}.
   \label{EOM-BG}
\eea
We can show
\bea
   w_\tau \equiv {p_\tau \over \mu_\tau}
       = { {\cal K} \over - {\cal K} + Q {\cal K}_{,Q} }, \quad
       c_\tau^2 \equiv {\dot p_\tau \over \dot \mu_\tau}
       = { {\cal K}_{,Q} \over Q {\cal K}_{,QQ} },
   \label{w-cs}
\eea
and
\bea
   \dot \mu_\tau + 3 H ( \mu_\tau + p_\tau ) = 0.
\eea

\subsection{Background constraint}
                                        \label{sec:BG-constraint}

For model-I in Eq.\ (\ref{model-I}), Eqs.\ (\ref{fluid-BG}), (\ref{EOM-BG}) and (\ref{w-cs}) give
\bea
   & & Q = 1 +
       \Big( {I_0 \over 2 \nu^2 {\cal K}_{n+1} a^3} \Big)^{1/n}, \quad
       \mu_\tau = {c^4 I_0 \over 8 \pi G a^3}
       {n Q + 1 \over n + 1},
   \nonumber \\
   & & w_\tau = {Q - 1 \over n Q + 1}, \quad
       c_\tau^2 = {1 \over n} \Big( 1 - {1 \over Q} \Big),
   \label{BG-sols}
\eea
where $a_0 = a (t_0) \equiv 1$ at present epoch. By a suitable parameter choice, with $[ I_0 / (2 \nu^2 {\cal K}_{n+1} a^3) ]^{1/n} \ll 1$, ${\cal K}$-term can simulate a dust fluid with $w_\tau \simeq 0 \simeq c_\tau^2$ in recent era. As $Q$ approaches $1$ the $\tau$-field behaves like a zero-pressure fluid to the background order with $I_0 = 8 \pi G \mu_{\tau0} a_0^3/c^4$. For $Q \neq 1$, $Q$ increases as $a$ decreases. Far in the early era with $I_0 / (2 \nu^2 {\cal K}_{n+1} a^3) \gg 1$, we have $w_\tau \simeq c_\tau^2 \simeq 1/n$ and $\mu_\tau \propto a^{-3(1+1/n)}$ \cite{Blanchet-Skordis-2024}.

For the dust-like behavior of $\tau$-field in the background, the authors of \cite{Blanchet-Skordis-2024} demand $w_{\tau*} \leq 0.0164$ at $a_* \sim 10^{-4.5}$. Using $I_0 \simeq 8 \pi G \mu_{\tau0}/c^4 = 3 \Omega_{\tau0} H_0^2/c^2$, we have
\bea
   {1 \over \nu \sqrt{{\cal K}_{n+1}}} \leq \Big[
       {2 c^2 \over 3 H_0^2 \Omega_{\tau0}}
       \Big( {(n+1) w_{\tau*} \over 1 - n w_{\tau*}} \Big)^n
       a_*^3 \Big]^{1/2} \equiv L_*.
   \label{constraint-BG}
\eea
For $n = 1$, with $\Omega_{\tau0} = 0.26$, the upper limit gives $L_* = 220 pc$. We have $L_* = 62$, $22$, $9.4$, $4.6$, and $0.57 pc$, respectively, for $n = 2, 3, 4, 5$, and $10$. This is a constraint on $\nu \sqrt{{\cal K}_{n+1}}$ parameter from the requirement of dust-like behavior of $\tau$-field in the background cosmology \cite{Blanchet-Skordis-2024}.

%
%
%
\section{PN approximation}
                                        \label{sec:PN}

The 1PN quantities required for Einstein's equation with a general fluid are presented in Appendix B of \cite{Hwang-Noh-axion-2023}. To 1PN order, the metric is
\bea
   & & g_{00} = - \Big( 1 + {2 \over c^2} \Phi \Big), \quad
       g_{0i} = - a {1 \over c^3} P_i,
   \nonumber \\
   & & g_{ij} = a^2 \Big( 1 - {2 \over c^2} \Psi \Big)
       \delta_{ij},
   \label{metric-PN}
\eea
where the index of $P_i$ is associated with $\delta_{ij}$ and its inverse. In this form we already imposed spatial gauge conditions \cite{Chandrasekhar-1965, Hwang-Noh-Puetzfeld-2008}. In the PN approximation, we use $x^0 = c t$. To 0PN order, $\Phi$ is the Newtonian gravitational potential; $\Psi$ and $P_i$ parts are 1PN order. To 1PN order, $\Phi$ contains 1PN order contribution as well. In Chandrasekhar's notation, we have \cite{Chandrasekhar-1965, Hwang-Noh-Puetzfeld-2008}
\bea
   \Phi = - U - {1 \over c^2}
       \left( 2 \overline \Phi - U^2 \right), \quad
       \Psi = - V.
   \label{Chandrasekhar-notation}
\eea
For the $\tau$-field, as an {\it ansatz}, we expand \cite{Flanagan-2023, Blanchet-Skordis-2024}
\bea
   \tau \equiv t + {1 \over c^2} \sigma
       + {1 \over c^4} \Sigma + \dots,
   \label{tau-PN}
\eea
where $\sigma$ and $\Sigma$ are functions of spacetime corresponding to 0PN and 1PN order, respectively. As we take $\tau = t$ to the background order, we have $Q = 1$ and $U_a = (-1, 0, 0, 0)$, thus ${\cal K} = 0$ and ${\cal J} = 0$, and the $\tau$-field does not contribute to the background order equations. Thus, our PN formulation is valid in the Friedmann background without ${\cal K}$ contribution, see below. Eq.\ (\ref{J-model-2}).

From Eq.\ (\ref{Q}), using the inverse metric in Eq.\ (B3) of \cite{Hwang-Noh-axion-2023}, we have
\bea
   & & \hskip -.8cm
       Q = 1 + {1 \over c^2} \Big( - \Phi + \dot \sigma
       - {1 \over 2 a^2} \sigma^{,i} \sigma_{,i} \Big)
   \nonumber \\
   & & \qquad
       \hskip -.8cm
       + {1 \over c^4} \Big\{ {3 \over 2} \Phi^2
       + \dot \Sigma - \Phi \dot \sigma
       + {1 \over a} P^i \sigma_{,i}
       - {1 \over a^2} \Big[
       {1 \over 8 a^2} ( \sigma^{,i} \sigma_{,i} )^2
   \nonumber \\
   & & \qquad
       \hskip -.8cm
       - {1 \over 2} \dot \sigma \sigma^{,i} \sigma_{,i}
       + \sigma^{,i} \Sigma_{,i}
       + \Big( \Psi + {1 \over 2} \Phi \Big)
       \sigma^{,i} \sigma_{,i} \Big] \Big\}.
   \label{Q-PN}
\eea
Apparently, the expansion is lengthy if we keep perturbations of $\tau$-field. In the following, we consider $\sigma$ to $0$PN order, and for simplicity, set $\sigma = 0 = \Sigma$ to $1$PN order. The meaning of $\sigma = 0 = \Sigma$ to 1PN order will be explained later.

\subsection{0PN limit}
                                        \label{sec:0PN}

To 0PN (Newtonian) order, we have
\bea
   Q = 1 - {1 \over c^2} \Xi, \quad
       \Xi \equiv \Phi - \dot \sigma
       + {1 \over 2 a^2} \sigma^{,i} \sigma_{,i}.
   \label{Xi-def}
\eea
The four-vector and acceleration in Eq.\ (\ref{Q}) give
\bea
   & & \hskip -1cm
       U_0 = - \Big[ 1 + {1 \over c^2} ( \Xi + \dot \sigma )
       \Big], \quad
       U_i = - {1 \over c} \sigma_{,i},
   \nonumber \\
   & & \hskip -1cm
       U^0 = 1 + {1 \over c^2} ( - 2 \Phi + \Xi
       + \dot \sigma ),
       \quad
       U^i = - {1 \over c} {1 \over a^2} \sigma^{,i};
   \nonumber \\
   & & \hskip -1cm
       A_i = \Xi_{,i}, \quad
       A_0 = {1 \over c} {1 \over a^2} \sigma^{,i} \Xi_{,i}, \quad
       A = {1 \over c^4} {1 \over a^2} \Xi^{,i} \Xi_{,i}.
\eea
For $\sigma = 0$, we have $U_a = n_a$.

As a concrete example, we consider the model-II in Eq.\ (\ref{model-II}). We have
\bea
   {\cal K} = {1 \over c^4} \nu^2 \Xi^2, \quad
       Q {\cal K}_{,Q} = - {2 \over c^2} \nu^2 \Xi.
\eea
To 0PN order the results are exactly the same as using the model-I in Eq.\ (\ref{model-I}) with $n = 1$. That is, {\it only} $n = 1$ contributes to a mass-like term in Eq.\ (\ref{Poisson-0PN}) to 0PN order \cite{Blanchet-Skordis-2024}. 

The observer's four-vector $u_a$ to 1PN order is presented in Eq.\ (B7) of \cite{Hwang-Noh-axion-2023}; to 0PN order we have $u_i = a v_i/c$ where the index of $v_i$ is associated with $\delta_{ij}$ and its inverse. We introduce
\bea
   q_i \equiv {a \over c} Q_i, \quad
       \pi_{ij} \equiv a^2 \Pi_{ij},
   \label{Q-Pi}
\eea
where indices of $Q_i$ and $\Pi_{ij}$ are associated with $\delta_{ij}$ and its inverse. Using Eq.\ (\ref{Tab}), Eq.\ (\ref{Tab-fluid}) for the $\tau$-field gives
\bea
   & & T_{00} = - {2 c^2 \over 8 \pi G}
       \Big[ {1 \over a^2}
       ( {\cal J}_{,A} \Xi^{,i} )_{,i} + \nu^2 \Xi \Big],
   \nonumber \\
   & & T_{0i} = - {2 c \over 8 \pi G}
       \Big[ {1 \over a^2}
       ( {\cal J}_{,A} \Xi^{,j} )_{,j} + \nu^2 \Xi \Big] \sigma_{,i},
   \nonumber \\
   & & T_{ij} = - {2 \over 8 \pi G}
       \Big\{ \Big[ {1 \over a^2}
       ( {\cal J}_{,A} \Xi^{,k} )_{,k} + \nu^2 \Xi \Big]
       \sigma_{,i} \sigma_{,j}
   \nonumber \\
   & & \qquad
       - {\cal J}_{,A} \Xi_{,i} \Xi_{,j}
       - {1 \over 2} a^2 ( \nu^2 \Xi^2 - {\cal J} c^4 )
       \delta_{ij} \Big\},
\eea
where ${\cal J} c^4$ is $c^0$-th order; as we set ${\cal J} = 0$ to the background order, ${\cal J} = \delta {\cal J} = {\cal J}_{,A} A$ and $A \sim {\cal O} (c^{-4})$. Using Eq.\ (\ref{fluid-quantities}), we have fluid quantities associated with the fluid four-vector $u_a$ of the $\tau$-fluid
\bea
   & & \hskip -.5cm
       \mu = - {2 c^2 \over 8 \pi G}
       \Big[ {1 \over a^2} ( {\cal J}_{,A} \Xi^{,i} )_{,i}
       + \nu^2 \Xi \Big],
   \nonumber \\
   & & \hskip -.5cm
       p = {1 \over 8 \pi G} {2 \over 3}
       \Big\{
       {1 \over a^2} {\cal J}_{,A} \Xi^{,i} \Xi_{,i}
       + {3 \over 2} ( \nu^2 \Xi^2 - {\cal J} c^4 )
   \nonumber \\
   & & \qquad
       \hskip -.5cm
       - \Big[ {1 \over a^2}
       ( {\cal J}_{,A} \Xi^{,k} )_{,k} + \nu^2 \Xi \Big]
       \Big( v^i + {1 \over a} \sigma^{,i} \Big)
       \Big( v_i + {1 \over a} \sigma_{,i} \Big)
       \Big\},
   \nonumber \\
   & & \hskip -.5cm
       Q_i = - \mu
       \Big( v_i + {1 \over a} \sigma_{,i} \Big),
   \nonumber \\
   & & \hskip -.5cm
       \Pi_{ij} = {1 \over 8 \pi G}
       \Big\{
       {2 \over a^2} {\cal J}_{,A} \Big( \Xi_{,i} \Xi_{,j}
       - {1 \over 3} \Xi^{,k} \Xi_{,k} \delta_{ij} \Big)
   \nonumber \\
   & & \qquad
       \hskip -.5cm
       - 2 \Big[ {1 \over a^2}
       ( {\cal J}_{,A} \Xi^{,k} )_{,k} + \nu^2 \Xi \Big]
       \Big[ \Big( v_i + {1 \over a} \sigma_{,i} \Big)
       \Big( v_j + {1 \over a} \sigma_{,j} \Big)
   \nonumber \\
   & & \qquad
       \hskip -.5cm
       - {1 \over 3} \Big( v^k + {1 \over a} \sigma^{,k} \Big)
       \Big( v_k + {1 \over a} \sigma_{,k} \Big)
       \delta_{ij} \Big] \Big\}.
\eea
To 0PN order, we have $\mu = \varrho c^2$.

As we have overlapping degrees of freedom in $u_a$ (velocity) and $q_a$ (flux), we can always take $q_a = 0$ as a frame choice, thus set $Q_i = 0$; this is often termed as the energy frame. Under this condition, we have
\bea
   v_i = - {1 \over a} \sigma_{,i}.
   \label{v-sigma}
\eea
The $\tau$-fluid is coupled {\it only} to the longitudinal part of the velocity. Fluid quantities become
\bea
   & & \varrho = - {2 \over 8 \pi G}
       \Big[ {1 \over a^2} ( {\cal J}_{,A} \Xi^{,i} )_{,i}
       + \nu^2 \Xi \Big],
   \nonumber \\
   & & p = {1 \over 8 \pi G}
       \Big( {2 \over 3 a^2}
       {\cal J}_{,A} \Xi^{,i} \Xi_{,i}
       + \nu^2 \Xi^2 - {\cal J} c^4 \Big),
   \nonumber \\
   & & \Pi_{ij} = {1 \over 8 \pi G}
       {2 \over a^2} {\cal J}_{,A} \Big( \Xi_{,i} \Xi_{,j}
       - {1 \over 3} \Xi^{,k} \Xi_{,k} \delta_{ij} \Big).
   \label{fluid-0PN}
\eea
The energy-momentum tensor of $\tau$-field becomes
\bea
   & & T^0_0 = - \varrho c^2, \quad
       T^0_i = - \varrho c \sigma_{,i}, \quad
       T^i_0 = {1 \over a^2} \varrho c \sigma^{,i},
   \nonumber \\
   & & T^i_j = p \delta^i_j + \Pi^i_j
       + {1 \over a^2} \varrho \sigma^{,i} \sigma_{,j}.
   \label{Tab-0PN}
\eea

The energy-momentum conservation, $T^b_{a;b} = 0$, for the $\tau$-field, gives
\bea
   & & {1 \over a^3} ( a^3 \varrho )^{\displaystyle{\cdot}}
       - {1 \over a^2} ( \varrho \sigma^{,i} )_{,i} = 0,
   \label{E-conserv-0PN} \\
   & & - \dot \sigma_{,i}
       + {1 \over a^2} \sigma^{,j} \sigma_{,ij}
       + \Phi_{,i}
       + {1 \over \varrho} ( p_{,i} + \Pi^j_{i,j} )
       = 0.
   \label{M-conserv-0PN}
\eea
Equation (\ref{E-conserv-0PN}) is the same as equation of motion with $S^0 = - 8 \pi G \varrho$ and $S^i = - S^0 \sigma^{,i}/(c a^2)$. By using $v_i$, the fluid velocity of $\tau$-field, we have
\bea
   & & {1 \over a^3} ( a^3 \varrho )^{\displaystyle{\cdot}}
       + {1 \over a} ( \varrho v^i )_{,i} = 0,
   \label{E-conserv-0PN-2} \\
   & & {1 \over a} ( a v_i )^{\displaystyle{\cdot}}
       + {1 \over a} v^j v_{i,j}
       + {1 \over a} \Phi_{,i}
       + {1 \over \varrho a} ( p_{,i} + \Pi^j_{i,j} )
       = 0.
   \label{M-conserv-0PN-2}
\eea
These are formally identical to conservation equations of an ordinary fluid with $p$ and $\Pi_{ij}$ given in Eq.\ (\ref{fluid-0PN}); for the baryon, we have $p = 0 = \Pi_{ij}$. We can show $p_{,i} + \Pi^j_{i,j} = - \varrho \Xi_{,i}$, and using Eq.\ (\ref{Xi-def}), Eq.\ (\ref{M-conserv-0PN-2}) is identically valid for the $\tau$-field.

To 0PN order, using curvature tensors presented in Eq.\ (B6) of \cite{Hwang-Noh-axion-2023}, and using Eq.\ (\ref{Tab-0PN}) we can derive Einstein's equation valid to 0PN order. To the background order, we have Eqs.\ (\ref{BG-eq1}) and (\ref{BG-eq2}). By subtracting the background equations, to 0PN order, $R^i_j$- and $R^0_0$-components of Einstein's equation, respectively, give $\Psi = \Phi$ and
\bea
   & & {\Delta \over a^2} \Phi
       = 4 \pi G ( \delta \varrho + \delta \varrho_\tau )
   \nonumber \\
   & & \qquad
       = 4 \pi G \delta \varrho
       - {1 \over a^2} ( {\cal J}_{,A} \Xi^{,i} )_{,i}
       - \nu^2 \Xi,
   \label{Poisson-0PN}
\eea
where $\delta \varrho$ is the perturbed density of ordinary fluids. This is Poisson's equation modified by $\tau$-field to 0PN order.

\subsubsection{Gauge transformation to 1PN order}
                                \label{sec:GT-PN}

Gauge transformation property in the PN approximation was studied in Sec.\ 6 of \cite{Hwang-Noh-Puetzfeld-2008}. Under a gauge (infinitesimal coordinate) transformation, $\widehat x^a = x^a + \xi^a$, we have
\bea
   \tau (x^e) = \widehat \tau (\widehat x^e)
       = \widehat \tau (x^e) + \tau_{,c} \xi^c.
   \label{tau-GT}
\eea
To 1PN order, as an {\it ansatz}, we consider
\bea
   \xi^0 = {1 \over c} \xi^{(2)0}
       + {1 \over c^3} \xi^{(4)0}, \quad
       \xi^i = {1 \over c^2} {1 \over a} \xi^{(2)i}.
\eea
Under a spatial gauge condition already imposed in our metric convention in Eq.\ (\ref{metric-PN}), we have $\xi^{(2)i} = 0$, thus $\xi^i = 0$ \cite{Hwang-Noh-Puetzfeld-2008}.

To 0PN and 1PN orders, respectively, we have
\bea
   \widehat \sigma = \sigma - \xi^{(2)0}, \quad
       \widehat \Sigma = \Sigma - \xi^{(4)0}
       - ( \sigma - \xi^{(2)0} ) \xi^{(2)0}.
\eea
From the gauge transformation of the metric $g_{0i}$, under our metric {\it ansatz} in Eq.\ (\ref{metric-PN}), we have $\xi^{(2)0} = \xi^{(2)0} (t)$ \cite{Hwang-Noh-Puetzfeld-2008}. The spatially constant $\xi^{(2)0}$ can be absorbed by a global redefinition of the time coordinate, and without losing generality we can set $\xi^{(2)0} \equiv 0$; this is necessary to maintain $V = U$, thus $\Psi = \Phi$ to 0PN order \cite{Hwang-Noh-Puetzfeld-2008}. Therefore, $\sigma$ {\it cannot} be used to fix the temporal gauge condition; in another words, to 0PN order, the PN approximation does not depend on the gauge which is natural as Newtonian gravity is free from the gauge issue \cite{Hwang-Noh-Puetzfeld-2008}. To 1PN order $\Sigma \equiv 0$ is a legitimate gauge condition which removes the gauge mode completely.

Still, we may set $\sigma = 0$, but it is a {\it physical} condition on the $\tau$-fluid \cite{Bonetti-2015}; it corresponds to setting $v_i = 0$ which is a serious condition on the fluid behavior of the $\tau$-field; $v_i = 0$ implies comoving with the $\tau$-fluid. The energy conservation equation in (\ref{E-conserv-0PN-2}) leads to $\varrho \propto a^{-3}$ for the $\tau$-fluid. To 1PN order, we expect $\Sigma = 0$ also implies comoving with the $\tau$-fluid, see $T^0_i$ in Eq.\ (\ref{Tab-1PN-q=0}). In perturbation theory later, $\sigma$ and $\Sigma$ can be used as the temporal gauge condition which removes the gauge degree of freedom completely; $\sigma \equiv 0 \equiv \Sigma$ corresponds to the comoving gauge of the $\tau$-fluid, see Sec.\ \ref{sec:GT-linear}. In our PN approximation, however, condition $\sigma = 0$ to 0PN order should be considered as a physical condition on the $\tau$-field.

\subsubsection{MOND regime with $\sigma = 0$}
                                        \label{sec:MOND-limit}

Equation (\ref{Poisson-0PN}) can be arranged as
\bea
   {1 \over a^2} \nabla \cdot \big( \nabla \Phi
       + {\cal J}_{,A} \nabla \Xi \big)
       + \nu^2 \Xi
       = 4 \pi G \delta \varrho.
\eea
For $\sigma = 0$, we have $\Xi = \Phi$, thus
\bea
   {1 \over a^2} \nabla \cdot \big[ ( 1 + {\cal J}_{,A} )
       \nabla \Phi \big]
       + \nu^2 \Phi
       = 4 \pi G \delta \varrho.
   \label{Poisson-MOND}
\eea
We have $A_i = \Phi_{,i}$, thus the acceleration becomes gravitational acceleration. A non-relativistic Lagrangian formulation of MOND by Bekenstein and Milgrom \cite{Bekenstein-Milgrom-1984}, demands
\bea
   1 + {\cal J}_{,A} = f ({1 \over a} |\nabla \Phi|/a_M) \equiv f(x),
   \label{1+J}
\eea
with $f(x) = x$ for $x \ll 1$ and $f(x) = 1$ for $x \gg 1$. We have
\bea
   A 
      = {1 \over c^4 a^2} \Phi^{,i} \Phi_{,i}
      = {1 \over c^4 a^2} |\nabla \Phi|^2
      = {a_M^2 \over c^4} x^2.
\eea
In order to achieve the MOND and the proper Newtonian gravity in the low and high acceleration regimes, respectively, we demand the following.

(i) In the MOND regime, $f(x) = x$ for $x \ll 1$, we need ${\cal J}_{,A} = x - 1 = c^2 \sqrt{A}/a_M - 1$, thus \cite{Blanchet-Skordis-2024}
\bea
   {\cal J} = - A + {2 \over 3} {c^2 \over a_M} A^{3/2}
      + {\cal O} (x^4).
   \label{J-MOND}
\eea

(ii) In the Newtonian regime, we need $f(x) = 1$ for $x \gg 1$. This implies ${\cal J}_{,A}$ should be negligibly small compared with unity in that limit \cite{Blanchet-Skordis-2024}.

For a simple interpolating function $f(x) = x/(1 + x)$, we have
\bea
   {\cal J}_{,A} = -{1 \over 1+x},
   \label{J-model}
\eea
and
\bea
   {\cal J} = - 2 {a_M \over c^2} \Big[ \sqrt{A}
       - {a_M \over c^2} \ln{\Big( 1
       + {c^2 \over a_M} \sqrt{A} \Big)} \Big].
   \label{J-model-2}
\eea

In galactic and smaller scales where MOND phenomena apply, the cosmic expansion effect is negligible and we may set $a \equiv 1$. In expanding cosmological background, as in Sec.\ \ref{sec:BG}, $\tau$-field with $\tau(t)$ contributes to the background evolution through ${\cal K}$. In our PN approximation, as we consider $\tau = t$ for the background, the $\tau$-field does not contribute to the background. For $\tau = \tau(t)$ in the background, it is {\it unclear} whether a proper PN expansion is possible. The requirement of a proper PN expansion may constrain allowed functional form of ${\cal K}$.

The MOND regime naturally appears by choosing ${\cal J}$-function in Eq.\ (\ref{J-MOND}). Still, we emphasize that setting $\sigma = 0$ is a physical condition imposed on the $\tau$-field.

\subsubsection{MOND constraint for $n = 1$}
                                        \label{sec:MOND-constraint}

The MOND limit is achieved by the ${\cal J}$-term. The ${\cal K}$-term is introduced only to account for the cosmological demands of the dark matter in the background and perturbations including the large-scale structure and CMB. The ${\cal K}$-term, however, introduces a mass-like correction term in the modified Poisson's equation, $\nu^2 \Phi$-term in Eq.\ (\ref{Poisson-MOND}) with $\nu = m_\nu c/\hbar$, which should be negligible. This term appears {\it only} for $n = 1$ in model-II in Eq.\ (\ref{model-II}). In order to recover phenomenology in galactic scale, we may need \cite{Blanchet-Skordis-2024}
\bea
   \nu^{-1} > 1 Mpc,
   \label{constraint-MOND}
\eea
corresponding to $m_\nu c^2 < 6.4 \times 10^{-30} {\rm eV}$. This constraint for a proper MOND phenomenology conflicts with the requirement for dust-like behavior in the background cosmology in Eq.\ (\ref{constraint-BG}), see \cite{Blanchet-Skordis-2024}. Thus, $n = 1$ part is {\it excluded}, and for $n \geq 2$ the model is free from this constraint.

\subsubsection{Faster growth in MOND regime}
                                        \label{sec:MOND-growth}

Here we show that the baryon density perturbation grows faster in the MOND regime compared with the Newtonian case. Equations (\ref{E-conserv-0PN-2}) and (\ref{M-conserv-0PN-2}) are valid for baryon with $p = 0 = \Pi_{ij}$. To the linear order, we have
\bea
   \dot \delta_b
       + {1 \over a} \nabla \cdot {\bf v}_b = 0, \quad
       {1 \over a} ( a {\bf v}_b )^{\displaystyle{\cdot}}
       = - {1 \over a} \nabla \Phi,
\eea
thus
\bea
   \ddot \delta_b + 2 H \dot \delta_b = {\Delta \over a^2} \Phi.
\eea
{\it Imposing} $\sigma = 0$, to the linear order, the MOND-modified Poisson's equation in Eq.\ (\ref{Poisson-MOND}) gives
\bea
   ( 1 + {\cal J}_{,A} ) {\Delta \over a^2} \Phi
       = 4 \pi G \delta \varrho_b,
\eea
where we ignored $\nu^2$-term from $n = 1$ as it is excluded in Eq.\ (\ref{constraint-MOND}). Combining these, we have
\bea
   \ddot \delta_b + 2 H \dot \delta_b
       = {4 \pi G \varrho_b \delta_b \over 1 + {\cal J}_{,A}}
       = \Big( 1 + {1 \over x} \Big)
       4 \pi G \varrho_b \delta_b,
   \label{delta-b-eq}
\eea
where in the second step we used the MOND function in Eq.\ (\ref{J-model}) with $x \equiv {1 \over a} |\nabla \Phi|/a_M$. In the MOND regime we have $x \ll 1$, and the gravity affecting density perturbation is stronger than in the Newtonian regime with $x \gg 1$.

For a pressureless background with $\Lambda = 0$, in the MOND regime we have a solution with ${1 \over a} \nabla \Phi = {1 \over a_0} \nabla \Phi_0$ and
\bea
   {1 \over a} | \nabla \Phi | =
       {3 \over 10} a_M \Omega_{b0}, \quad
       \delta_b = {{3 \over 10} \Omega_{b0} {1 \over a} \Delta \Phi
       \over 4 \pi G \varrho_b a^3} a^2.
\eea
Thus, $\delta_b \propto a^2$ grows faster than in the CDM case where $\delta_b \propto a$.

The faster growth of galaxies was previously noticed in the MOND paradigm \cite{Sanders-1998, McGaugh-2024}; in the low-acceleration MOND regime, we have $|\vec{a}| = \sqrt{ a_M |\vec{g}_N| } > |\vec{g}_N|$, thus the gravity is stronger than Newtonian gravity, and the structures can grow faster. Here, we showed the case using the BMS model and our result agrees with a heuristic derivation in \cite{Nusser-2002}.

\subsection{1PN approximation with $\sigma = 0 = \Sigma$}
                                        \label{sec:1PN}

To 1PN order, for simplicity we set $\sigma = 0 = \Sigma$; $\sigma = 0$ is a physical condition we impose whereas $\Sigma = 0$ is a temporal gauge condition we choose. From Eq.\ (\ref{Q-PN}), we have
\bea
   Q = 1 - {1 \over c^2} \Phi + {3 \over 2 c^4} \Phi^2.
   \label{Q-1PN}
\eea
Equation (\ref{Q}) gives
\begin{widetext}
\bea
   & & U_0 = - \Big( 1 + {1 \over c^2} \Phi
       - {1 \over 2 c^4} \Phi^2 \Big), \quad
       U_i = 0, \quad
       U^0 = 1 - {1 \over c^2} \Phi
       + {3 \over 2 c^4} \Phi^2, \quad
       U^i = {1 \over c^3} {1 \over a} P^i; \quad
       A_0 = - {1 \over c^3} {1 \over a} P^i \Phi_{,i},
   \nonumber \\
   & & A_i = \Phi_{,i} \Big( 1 - {2 \over c^2} \Phi \Big), \quad
       A^0 = 0, \quad
       A^i = {1 \over a^2} \Phi^{,i} \Big[ 1
       + {2 \over c^2} ( \Psi - \Phi ) \Big], \quad
       A = {1 \over c^4} {1 \over a^2} \Phi^{,i} \Phi_{,i}
       \Big[ 1 + {2 \over c^2} ( \Psi - 2 \Phi ) \Big].
\eea
As we set $\sigma = 0 = \Sigma$, $U_a$ is the same as the normal four-vector. By taking the model-II in Eq.\ (\ref{model-II}), we have
\bea
   {\cal K} = {1 \over c^4} \nu^2 \Phi^2
       \Big( 1 - {3 \over c^2} \Phi \Big), \quad
       Q {\cal K}_{,Q} = - {2 \over c^2} \nu^2 \Phi
       \Big[ 1 - {1 \over c^2}
       \Big( {5 \over 2} + {\cal K}_3 \Big) \Phi \Big].
\eea
Thus, only $n = 1$ and $2$ contribute to 1PN order.

Using Eq.\ (\ref{Tab}), Eq.\ (\ref{Tab-fluid}) for the $\tau$-field gives
\bea
   & & T_{00} = {c^2 \over 8 \pi G}
       \Big[ - {2 \over a^2} ( {\cal J}_{,A} \Phi^{,i} )_{,i}
       \Big( 1 + {2 \over c^2} \Psi \Big)
       + {2 \over a^2 c^2} {\cal J}_{,A}
       ( \Phi + \Psi )_{,i} \Phi^{,i}
       - 2 \nu^2 \Phi \Big( 1
       - {1 \over c^2} {\cal K}_3 \Phi \Big)
       + {\cal J} c^2 \Big], \quad
       T_{0i} = 0,
   \nonumber \\
   & & T_{ij} = {1 \over 8 \pi G} \Big\{ a^2 \Big[ \nu^2 \Phi^2
       \Big( 1 - {3 \over c^2} \Phi - {2 \over c^2} \Psi \Big)
       - {\cal J} c^4 \Big( 1 - {2 \over c^2} \Psi \Big) \Big]
       \delta_{ij}
       + 2 {\cal J}_{,A} \Phi_{,i} \Phi_{,j}
       \Big( 1 - {4 \over c^2} \Phi \Big) \Big\}.
\eea
Using Eqs.\ (\ref{Tab}), (\ref{fluid-quantities}) and (\ref{Q-Pi}), the $\tau$-fluid quantities are
\bea
   & & \hskip - .5cm
       \mu = {1 \over 8 \pi G} \Big\{
       - {2 \over a^2} ( {\cal J}_{,A} \Phi^{,i} )_{,i}
       ( c^2 + v^2 - 2 \Phi + 2 \Psi )
       + {2 \over a^2} {\cal J}_{,A} ( \Phi + \Psi )^{,i} \Phi_{,i}
       - 2 \nu^2 \Phi \big[ c^2 + v^2
       - ( 2 + {\cal K}_3 ) \Phi \big]
       + {\cal J} c^4 \Big\},
   \nonumber \\
   & & \hskip - .5cm
       p = {1 \over 8 \pi G} {1 \over 3} \Big\{
       - {2 \over a^2} ( {\cal J}_{,A} \Phi^{,i} )_{,i}
       \Big[ 1 + {1 \over c^2} ( v^2 - 2 \Phi + 4 \Psi )
       \Big] v^2
       - 2 \nu^2 \Phi \Big[ 1 + {1 \over c^2}
       \Big( v^2 - {5 \over 2} \Phi - {\cal K}_3 \Phi
       + 2 \Psi \Big) \Big] v^2
       + 3 \nu^2 \Phi^2 \Big( 1 - {3 \over c^2} \Phi \Big)
   \nonumber \\
   & & \qquad
       \hskip - .5cm
       + {2 \over a^2} {\cal J}_{,A} \Phi^{,i} \Phi_{,i}
       \Big[ 1 + {1 \over c^2} ( 2 \Psi - 4 \Phi ) \Big]
       + {2 \over a^2 c^2} {\cal J}_{,A}
       (\Phi + \Psi )_{,i} \Phi^{,i} v^2
       + {2 \over a^2 c^2} {\cal J}_{,A} \Phi_{,i} v^i
       \Phi_{,j} v^j
       - 3 {\cal J} c^4 \Big\},
   \nonumber \\
   & & \hskip - .5cm
       Q_i = {1 \over 8 \pi G} \Big\{
       {2 \over a^2} v_i ( {\cal J}_{,A} \Phi^{,j} )_{,j}
       \Big( c^2 + {3 \over 2} v^2 - 2 \Phi + 2 \Psi \Big)
       - {2 \over a^2} {\cal J}_{,A}
       \big[ \Phi_{,i} \Phi^{,j} v_j
       + v_i (\Phi + \Psi)^{,j} \Phi_{,j} \big]
   \nonumber \\
   & & \qquad \hskip - .5cm
       + \nu^2 \Phi v_i \big[ 2 c^2 + 3 v^2
       - ( 5 + 2 {\cal K}_3 ) \Phi \big] \Big\},
   \nonumber \\
   & & \hskip - .5cm
       \Pi_{ij} = {1 \over 8 \pi G} \Big\{
       - \Big( v_i v_j - {1 \over 3} \delta_{ij} v^2 \Big)
       \Big[ {2 \over a^2} ( {\cal J}_{,A} \Phi^{,k} )_{,k}
       \Big( 1 - {2 \over c^2} \Phi + {2 \over c^2} \Psi
       + {1 \over c^2} v^2 \Big)
       + 2 \nu^2 \Phi \Big( 1 + {1 \over c^2} v^2
       - {5 \over 2 c^2} \Phi
       - {1 \over c^2} {\cal K}_3 \Phi \Big)
   \nonumber \\
   & & \qquad \hskip - .5cm
       - {2 \over a^2 c^2} {\cal J}_{,A}
       ( \Phi + \Psi )_{,k} \Phi^{,k} \Big]
       + {2 \over a^2} {\cal J}_{,A}
       \Big( \Phi_{,i} \Phi_{,j}
       - {1 \over 3} \delta_{ij} \Phi^{,k} \Phi_{,k} \Big)
       \Big( 1 - {4 \over c^2} \Phi \Big)
       + {2 \over a^2 c^2} {\cal J}_{,A} \Phi_{,k} v^k
       \Big( v_i \Phi_{,j} + v_j \Phi_{,i}
       - {2 \over 3} \delta_{ij} \Phi_{,\ell} v^\ell \Big)
   \nonumber \\
   & & \qquad \hskip - .5cm
       - {2 \over 3 c^2} v_i v_j \Big[ {2 \over a^2}
       ( {\cal J}_{,A} \Phi^{,k} )_{,k} v^2
       + 2 \nu^2 \Phi v^2
       + {1 \over a^2} {\cal J}_{,A} \Phi^{,k} \Phi_{,k} \Big]
       + {2 \over 3 c^2 a^2} \delta_{ij}
       {\cal J}_{,A} \Phi_{,k} v^k \Phi_{,\ell} v^\ell
       \Big\}.
\eea
To 1PN order, we have $\mu = \varrho ( c^2 + \Pi )$ with $\varrho \Pi$ the internal energy density \cite{Chandrasekhar-1965} and $\Pi^i_i = {1 \over c^2} \Pi_{ij} v^i v^j$. The energy-frame condition $q_i \equiv 0$ for the $\tau$-fluid implies $v_i = 0$, thus the velocity four-vector becomes normal, $u_a = n_a$. The fluid quantities become
\bea
   & & \mu = {1 \over 8 \pi G} \Big\{
       - {2 \over a^2} ( {\cal J}_{,A} \Phi^{,i} )_{,i}
       ( c^2 - 2 \Phi + 2 \Psi )
       + {2 \over a^2} {\cal J}_{,A} ( \Phi + \Psi )^{,i} \Phi_{,i}
       - 2 \nu^2 \Phi \big[ c^2
       - ( 2 + {\cal K}_3 ) \Phi \big]
       + {\cal J} c^4 \Big\},
   \nonumber \\
   & & p = {1 \over 8 \pi G} \Big[
       \nu^2 \Phi^2 \Big( 1 - {3 \over c^2} \Phi \Big)
       + {2 \over 3 a^2} {\cal J}_{,A} \Phi^{,i} \Phi_{,i}
       \Big( 1 - {4 \over c^2} \Phi+ {2 \over c^2} \Psi \Big)
       - {\cal J} c^4
       \Big],
   \nonumber \\
   & & \Pi_{ij} = {1 \over 8 \pi G}
       {2 \over a^2} {\cal J}_{,A}
       \Big( \Phi_{,i} \Phi_{,j}
       - {1 \over 3} \delta_{ij} \Phi^{,k} \Phi_{,k} \Big)
       \Big( 1 - {4 \over c^2} \Phi \Big).
   \label{fluid-1PN-q=0}
\eea
For the $\tau$-fluid, using Eq.\ (\ref{fluid-1PN-q=0}), and Eq.\ (B9) in \cite{Hwang-Noh-axion-2023}, we can show
\bea
   T^0_0 = - \varrho c^2
       \Big( 1 + {1 \over c^2} \Pi \Big), \quad
       T^0_i = 0, \quad
       T^i_0 = - {1 \over c a} \varrho P^i, \quad
       T^i_j = p \delta^i_j
       + \Big( 1 + {2 \over c^2} \Psi \Big) \Pi^i_j.
   \label{Tab-1PN-q=0}
\eea

We note that setting $q_i = 0$ is a frame choice always allowed for any energy-momentum component. Under this choice, the flux or momentum of the component is encoded in the fluid four-vector $u_a$. As the four-vector becomes normal with $n_i \equiv 0$, we have the physical momentum of the component vanish; this is reflected in $T^0_i = (\mu + p) u^0 u_i = 0$ to 1PN order. This implies that the $\tau$-fluid quantities are measured by an observer comoving with the $\tau$-fluid. This happened as we imposed $\sigma = 0 = \Sigma$ as the physical and gauge conditions.

Using Eq.\ (\ref{Tab-1PN-q=0}), $T^b_{a;b} = 0$ for $\tau$-fluid gives
\bea
   & & \dot \mu + 3 H ( \mu + p )
       - 3 \varrho \dot \Psi
       + {1 \over a} ( \varrho P^i )_{,i} = 0,
   \label{E-conserv-1PN} \\
   & & \varrho \Phi_{,i} + p_{,i} + \Pi^j_{i,j}
       + {1 \over c^2} \big\{
       \Phi_{,i} [ \varrho ( \Pi - 2 \Phi ) + p ]
       + ( \Phi - \Psi )_{,j} \Pi^j_i
       + 2 \Psi \Pi^j_{i,j} \big\} = 0.
   \label{mom-conserv-1PN}
\eea
These coincide with the energy and momentum conservation equations of a fluid in (B23) and (B24) of \cite{Hwang-Noh-axion-2023} by setting $v_i = 0 = Q_i$. For a fluid, $Q_i = 0$ can be regarded as the frame choice, and $v_i = 0$ implies that the fluid velocity vanishes which can be achieved {\it in} relativistic perturbation theory by a temporal gauge choice; the comoving gauge condition for a fluid imposes only the longitudinal part of $v_i$ to vanish. In our $\tau$-fluid case, $\sigma = 0 = \Sigma$ implies $v_i = 0$, thus the transverse part also vanishes; in the perturbation theory in next section we will show that the $\tau$-field does not support the (transverse) vector-type perturbation.

Equation of motion in Eqs.\ (\ref{EOM}) and (\ref{Sa}) gives
\bea
   & & \dot S^0 + \Big[ 3 H + {1 \over c^2}
       ( \dot \Phi - 3 \dot \Psi ) \Big] S^0
       + c S^i_{\;\;,i} = 0,
   \label{EOM-1PN} \\
   & & S^0 = {1 \over c^2} ( - 8 \pi G \mu + c^4 {\cal J}
       - \mu^2 \Phi^2 ), \quad
       S^i = - {1 \over c^3} \Big(
       {2 \over a^2} {\cal J}_{,A} \dot \Phi \Phi^{,i}
       + 8 \pi G \varrho {1 \over a} P^i \Big),
\eea
and it gives Eq.\ (\ref{E-conserv-1PN}).

Using the curvature tensors presented in Eq.\ (B6) of \cite{Hwang-Noh-axion-2023}, we can derive Einstein's equation valid to 1PN order. To the background order, we have Eqs.\ (\ref{BG-eq1}) and (\ref{BG-eq2}). By subtracting the background equations, to 1PN order, $R^0_0$-, $R^i_j$- and $R^0_i$-components of Einstein's equation, respectively, give
\bea
   & & \hskip -.7cm
       {\Delta \over a^2} \Phi
       + {1 \over c^2} \Big\{ 3 \Big[ \ddot \Psi
       + H ( \dot \Phi + 2 \dot \Psi )
       + 2 {\ddot a \over a} \Phi \Big]
       - {1 \over a^2} \big[ 2 ( \Phi - \Psi ) \Delta \Phi
       + (\Phi + \Psi )^{,i} \Phi_{,i} \big]
       - {1 \over a^2} ( a P^i_{\;\;,i} )^{\displaystyle{\cdot}} \Big\}
        +\nu^2 \Phi \Big[ 1
        - {1 \over c^2} \Big( {7 \over 2}
        + {\cal K}_3 \Big) \Phi \Big]
   \nonumber \\
   & & \qquad
       \hskip -.7cm
       = 4 \pi G \Big\{ \delta \Big[ \varrho
       \Big( 1 + {1 \over c^2} \Pi \Big) \Big]
       + {3 \over c^2} \delta p
       + {2\over c^2} \varrho v^2 \Big\}
       - {1 \over a^2} ( {\cal J}_{,A} \Phi^{,i} )_{,i}
       \Big[ 1 + {2 \over c^2} ( \Psi - \Phi ) \Big]
       - {\cal J} c^2
       + {1 \over a^2 c^2} {\cal J}_{,A}
       ( 2 \Phi + \Psi )_{,i} \Phi^{,i},
   \label{1PN-1} \\
   & & \hskip -.7cm
       {\Delta \over a^2} \Psi \delta^i_j
       + {1 \over a^2} ( \Psi - \Phi )^{,i}_{\;\;\;j}
       = \Big[ 4 \pi G \delta \varrho
       - {1 \over a^2} ( {\cal J}_{,A} \Phi^{,k} )_{,k}
       - \nu^2 \Phi \Big] \delta^i_j,
   \label{1PN-2} \\
   & & \hskip -.7cm
       - 2 ( \dot \Psi + H \Phi )_{,i}
       + {1 \over 2 a} ( P^j_{\;\;,ji} - \Delta P_i )
       = 8 \pi G \varrho a v_i.
   \label{1PN-3}
\eea
\end{widetext}
Equation (\ref{1PN-2}) is valid to 0PN order whereas the other two equations are valid to 1PN order. To 0PN order, Eq.\ (\ref{1PN-1}) gives Eq.\ (\ref{Poisson-0PN}) with $\sigma = 0$. To 0PN order, Eq.\ (\ref{1PN-2}) is the same as Eq.\ (\ref{1PN-1}), with $\Psi = \Phi$.

The fluid quantities in Eqs.\ (\ref{1PN-1})-(\ref{1PN-3}) are the sum of other fluids besides the $\tau$-fluid; we may include the $\tau$-field to the collective fluid quantities as $\mu = \mu_m + \mu_\tau$, $p = p_m + p_\tau$, and $\varrho v_i = \varrho^m v^m_i + \varrho^\tau v^\tau_i$. For the $\tau$-fluid we have conservation equations in (\ref{E-conserv-1PN})-(\ref{mom-conserv-1PN}) or the equation of motion in Eq.\ (\ref{EOM-1PN}). The fluid conservation equations to 1PN order are in Eqs.\ (B23) and (B24) of \cite{Hwang-Noh-axion-2023}. These are the complete set of equations valid to 1PN order with $\Sigma = 0$ as the temporal gauge condition; for other choice of gauge conditions, see Sec.\ 6 in \cite{Hwang-Noh-Puetzfeld-2008}.

%
%
%
\section{Perturbation theory}
                                        \label{sec:PT}

We consider a metric
\bea
   & & g_{00} = - a^2 ( 1 + 2 \alpha ), \quad
       g_{0i} = - a \chi_i,
   \nonumber \\
   & & g_{ij} = a^2 ( 1 + 2 \varphi ) \delta_{ij},
   \label{metric-FNLE}
\eea
where $\alpha$, $\varphi$ and $\chi_i$ are functions of spacetime with arbitrary amplitudes and the index of $\chi_i$ is associated with $\delta_{ij}$ and its inverse; here, we have $x^0 = \eta$. In this form, we ignored the tracefree-transverse tensor type perturbation, and imposed a spatial gauge condition which (together with appropriate choice of temporal gauge condition) removes the gauge degrees of freedom completely. In \cite{Hwang-Noh-2013} we derived the fully nonlinear and exact perturbation equations, and the case is extended to more general metric including the tensor type peturbation and without imposing spatial gauge condition in \cite{Gong-2017}. The fully nonlinear and exact formulation useful for our analysis is summarized in Appendix C of \cite{Hwang-Noh-EM-2023}.

We set
\bea
   \tau \equiv \bar \tau (t)
       + {1 \over c^2} \sigma ({\bf x}, t),
   \label{tau-FNLE}
\eea
where $\sigma$ is a function of spacetime with arbitrary amplitude; in PN expansion $\Sigma$ and other higher order expansions in Eq.\ (\ref{tau-PN}) are absorbed to $\sigma$; if needed we use an overbar to indicate the background order quantity. The fully nonlinear and exact perturbation formulation includes the PN expansion; comparing Eqs.\ (\ref{metric-FNLE}) and (\ref{metric-PN}), we have
\bea
   \alpha = {\Phi \over c^2}, \quad
       \varphi = - {\Psi \over c^2}, \quad
       \chi_i = a {P_i \over c^3}.
\eea

Equation (\ref{Q}), using the exact inverse metric in Eq.\ (C3) of \cite{Hwang-Noh-EM-2023}, we have
\bea
   & & \hskip -.9cm
       Q = {1 \over {\cal N}} \Big[
       \Big( \dot {\bar \tau} + {1 \over c^2} \dot \sigma \Big)^2
       + {2 \chi^i \over a^2 ( 1 + 2 \varphi )}
       \Big( \dot {\bar \tau} + {1 \over c^2} \dot \sigma \Big)
       {1 \over c} \sigma_{,i}
   \nonumber \\
   & & \qquad
       \hskip -.9cm
       - {1 \over a^2 ( 1 + 2 \varphi )}
       \Big( {\cal N}^2 \delta^{ij}
       - {\chi^i \chi^j \over a^2 ( 1 + 2 \varphi )} \Big)
       {1 \over c^2} \sigma_{,i} \sigma_{,j} \Big]^{1/2},
   \label{Q-perturbation}
\eea
with ${\cal N}$ given in Eq.\ (\ref{Q-FNLE}). Apparently, the expansion is lengthy if we keep perturbations of $\sigma$. In the following, we consider $\sigma$ in the linear order perturbation theory, and for simplicity, set $\sigma = 0$ in the fully nonlinear and exact formulation; $\sigma = 0$ corresponds to a temporal gauge condition valid to fully nonlinear order and is the same as the comoving gauge of the $\tau$-fluid, see Eqs.\ (\ref{fluid-linear}) and (\ref{fluid-FNLE}).

\subsection{Linear perturbation}
                                        \label{sec:linear}

To the linear order in perturbations, Eq.\ (\ref{Q-perturbation}) gives
\bea
   Q = ( 1 - \alpha ) \bar Q + {1 \over c^2} \dot \sigma, \quad
       \bar Q = \dot {\bar \tau}.
   \label{Q-linear}
\eea
The four-vector and acceleration become
\bea
   & & U_0 = - a ( 1 + \alpha ), \quad
       U_i = - {1 \over c Q} \sigma_{,i},
   \nonumber \\
   & & U^0 = {1 \over a} ( 1 - \alpha ), \quad
       U^i = {1 \over a^2} \Big( \chi^i
       - {1 \over c Q} \sigma^{,i} \Big);
   \nonumber \\
   & & A_0 = 0, \quad
       A_i = c^2 \Upsilon_{,i}, \quad
       A^0 = 0, \quad
       A^i = {c^2 \over a^2} \Upsilon^{,i},
   \nonumber \\
   & & \Upsilon \equiv \alpha - {1 \over c^2 Q}
       \Big( \dot \sigma - {\dot Q \over Q} \sigma \Big)
       \equiv \alpha_\sigma, \quad
       A = 0,
   \label{Upsilon}
\eea
where $\alpha_\sigma$ is a unique gauge-invariant notation which becomes $\alpha$ in the $\sigma = 0$ gauge condition, see later. To the linear order, we have $A = 0$. However, in the MOND regime, from Eq.\ (\ref{J-MOND}), we need ${\cal J}_{,A} = - 1 + (ac^2/a_M) \sqrt{A}$ and $\sqrt{A}$ is linear order. We have $\sqrt{A} = (\Upsilon^{,i} \Upsilon_{,i})^{1/2}/a = |\nabla \Upsilon|/a$, thus ${\cal J}_{,A} = - 1 + |\nabla \Upsilon| c^2/a_M$ which is consistent with Eq.\ (\ref{1+J}) in the MOND regime by identifying $\Upsilon = \alpha_\sigma = \Phi/c^2$. In Newtonian regime, we have ${\cal J}_{,A} = 0$.

The $\tau$-field energy-momentum tensor becomes
\bea
   & & \hskip -.5cm
       T_{00}
       = {c^4 \over 8 \pi G} a^2 \Big[ ( 1 + 2 \alpha ) ( {\cal J} - {\cal K} + Q {\cal K}_{,Q} )
       - 2 {\cal J}_{,A} {\Delta \over a^2} \Upsilon \Big],
   \nonumber \\
   & & \hskip -.5cm
       T_{0i}
       = {c^4 \over 8 \pi G} a \Big[ ( {\cal J} - {\cal K} )
       \chi_i
       + {\cal K}_{,Q} {1 \over c} \sigma_{,i} \Big],
   \nonumber \\
   & & \hskip -.5cm
       T_{ij}
       = {c^4 \over 8 \pi G} a^2 ( 1 + 2 \varphi )
       ( - {\cal J} + {\cal K} ) \delta_{ij}.
\eea
The ADM fluid quantities are
\bea
   E \equiv n_a n_b T^{ab},
       \quad
       J_i \equiv - n_b T^b_i, \quad
       S_{ij} \equiv T_{ij}.
   \label{ADM-fluid-def}
\eea
where indices of  $J_i$ and $S_{ij}$ are associated with the intrinsic ADM metric, $\overline h_{ij} (\equiv g_{ij})$, and its inverse. Using  Eqs.\ (117) and (C1) of \cite{Hwang-Noh-EM-2023} we have
\bea
   E \equiv \mu, \quad
       J_i \equiv a c m_i, \quad
       S_{ij} \equiv a^2 ( 1 + 2 \varphi ) m_{ij},
   \label{fluid-ADM-1}
\eea
where indices of $m_i$ and $m_{ij}$ are associated with $\delta_{ij}$ and its inverse. Using
\bea
   & & m_i = {1 \over c^2} ( \mu + p ) v_i, \quad
       v_i \equiv - v_{,i} + v^{(v)}_i, \quad
       v^{(v)i}_{\;\;\;\;\;\;,i} \equiv 0,
   \nonumber \\
   & & p = {1 \over 3} m^k_k, \quad
       \overline m_{ij} \equiv m_{ij} - {1 \over 3} m^k_k \delta_{ij},
   \label{fluid-ADM-2}
\eea
we have the $\tau$-fluid quantities
\bea
   & & \mu = {c^4 \over 8 \pi G}
       \Big( {\cal J} - {\cal K} + Q {\cal K}_{,Q}
       - 2 {\cal J}_{,A} {\Delta \over a^2} \Upsilon \Big),
   \nonumber \\
   & & ( \mu + p ) v
       = {c^4 \over 8 \pi G} {1 \over a} {\cal K}_{,Q} \sigma, \quad
       v^{(v)}_i = 0,
   \nonumber \\
   & & p = {c^4 \over 8 \pi G} ( - {\cal J} + {\cal K} ), \quad
       \overline m_{ij} = 0.
   \label{fluid-linear}
\eea
To the background order, we recover Eq.\ (\ref{fluid-BG}). The perturbed parts are
\bea
   & & \delta \mu = {c^4 \over 8 \pi G}
       \Big( Q {\cal K}_{,QQ} \delta Q
       - 2 {\cal J}_{,A} {\Delta \over a^2} \Upsilon \Big),
   \nonumber \\
   & & \delta p = {c^4 \over 8 \pi G} {\cal K}_{,Q} \delta Q, \
       \quad
       v = {1 \over a Q} \sigma.
   \label{fluid-pert}
\eea
Only scalar-type perturbations are excited without anisotropic stress. We have
\bea
   {\delta p \over \delta \mu}
       = c_\tau^2 \Big( 1 - {2 c_\tau^2 \over
       {\cal K}_{,Q} \delta Q}
       {\cal J}_{,A} {\Delta \over a^2} \Upsilon \Big)^{-1}.
   \label{delta-p-mu}
\eea
The entropic perturbation (isotropic stress) of $\tau$-fluid is
\bea
   e \equiv \delta p - c_\tau^2 \delta \mu
      = {c^4 \over 8 \pi G} 2 c_\tau^2 {\cal J}_{,A}
      {\Delta \over a^2} \Upsilon.
   \label{e-linear}
\eea
As we have $A = 0$ and ignore the constant part of ${\cal J}$, we have ${\cal J} = 0$ whereas ${\cal J}_{,A} \neq 0$. We note that for vanishing ${\cal K}$, we have $p_\tau = 0 = c_\tau^2$, thus $e = 0$.

Einstein's equations in Eqs.\ (C8)-(C12) of \cite{Hwang-Noh-EM-2023} give
\bea
   & & \hskip -.8cm
       \kappa \equiv 3 H \alpha - 3 \dot \varphi
       - c {\Delta \over a^2} \chi,
   \label{eq1-linear} \\
   & & \hskip -.8cm
       H \kappa + c^2 {\Delta \over a^2} \varphi
       = - {4 \pi G \over c^2} \delta \mu,
   \label{eq2-linear} \\
   & & \hskip -.8cm
       \kappa + c {\Delta \over a^2} \chi
       = {12 \pi G \over c^4} a (\mu + p) v,
   \label{eq3-linear} \\
   & & \hskip -.8cm
       \dot \kappa + 2 H \kappa
       + \Big( c^2 {\Delta \over a^2} + 3 \dot H \Big) \alpha
       = {4 \pi G \over c^2} ( \delta \mu + 3 \delta p ),
   \label{eq4-linear} \\
   & & \hskip -.8cm
       c ( \dot \chi + H \chi ) - c^2 (\alpha + \varphi ) = 0,
   \label{eq5-linear}
\eea
where the fluid quantities are the collective ones, for example, $\delta \mu = \delta \mu_m + \delta \mu_\tau$, $\delta p = \delta p_m + \delta p_\tau$, $(\mu + p) v = (\mu_m + p_m) v_m + (\mu_\tau + p_\tau) v_\tau$; the matter part include baryon, photon, neutrino, etc.; we decomposed $\chi_i \equiv \chi_{,i} + \chi_i^{(v)}$ with $\chi^{(v)i}_{\;\;\;\;\;\; , i} \equiv 0$. We ignored the anisotropic stress of ordinary fluid in Eq.\ (\ref{eq5-linear}). The vector- and tensor-type perturbation equations are {\it not} affected by the $\tau$-field.

The equation of motion in Eq.\ (\ref{EOM}) gives
\bea
   & & {\Delta \over a} ( a {\cal J}_{,A} \Upsilon
       )^{\displaystyle{\cdot}}
       = {\bar Q \over 2 a} \big\{ a^3 \big[
       ( 1 + 3 \varphi ) \bar {\cal K}_{,Q}
       + \delta ({\cal K}_{,Q} ) \big] \big\}^{\displaystyle{\cdot}}
   \nonumber \\
   & & \qquad
       + {c \over 2} Q {\cal K}_{,Q}
       \Delta \Big( \chi - {1 \over c Q} \sigma \Big).
   \label{EOM-linear}
\eea
To background order, we recover Eq.\ (\ref{EOM-BG}).

The energy and momentum conservation equations in Eqs.\ (C14) of \cite{Hwang-Noh-EM-2023} give
\bea
   & & \hskip -1.cm
       \delta \dot \mu + 3 H ( \delta \mu + \delta p )
       = ( \mu + p ) \Big( \kappa - 3 H \alpha
       + {\Delta \over a} v \Big),
   \label{E-conserv-linear} \\
   & & \hskip -1.cm
       {1 \over a^4} [ a^4 ( \mu + p ) v
       ]^{\displaystyle{\cdot}}
       = {c^2 \over a} [ ( \mu + p ) \alpha + \delta p ].
   \label{mom-conserv-linear}
\eea
These two equations are valid for ordinary fluid with the anisotropic stress ignored in the second equation; we have $p = \delta p = 0$ for dust and $p = {1 \over 3} \mu$ and $\delta p = {1 \over 3} \delta \mu$ for photon fluid, etc. These are also valid for the $\tau$-fluid with the $\tau$-fluid quantities given in Eq.\ (\ref{fluid-linear}). Using the $\tau$-fluid quantities, Eq.\ (\ref{E-conserv-linear}) is naturally valid by the equation of motion in Eq.\ (\ref{EOM-linear}) together with Einstein's equation and $\dot {\cal J} = 0$. Using Eqs.\ (\ref{EOM-BG}), (\ref{Q-linear}) and (\ref{fluid-linear}), Eq.\ (\ref{mom-conserv-linear}) is identically valid.

The complete set of linear perturbation equations without imposing temporal gauge condition is Eqs.\ (\ref{eq1-linear})-(\ref{eq5-linear}) from Einstein's equation, Eqs.\ (\ref{E-conserv-linear}) and (\ref{mom-conserv-linear}) for other fluids, and the same equation for the $\tau$-fluid with the fluid quantities in Eq.\ (\ref{fluid-linear}). For the $\tau$-fluid, instead of conservation equations in (\ref{E-conserv-linear}) and (\ref{mom-conserv-linear}), we can directly use the field variable and the equation of motion in Eq.\ (\ref{EOM-linear}).

\subsubsection{Gauge transformation}
                                 \label{sec:GT-linear}

Under the gauge transformation, $\widehat x^a = x^a + \xi^a$, using Eq.\ (\ref{tau-GT}), Eqs.\ (\ref{tau-FNLE}) and (\ref{Q-linear}) give
\bea
   & & {\widehat \sigma \over c^2}
       = {\sigma \over c^2}
       - \tau^\prime \xi^0, \quad
       \widehat Q = Q - c \Big( {\tau^\prime \over a}
       \Big)^\prime \xi^0,
   \nonumber \\
   & & \widehat {\cal K} = {\cal K}
       - {\cal K}_{,Q} c \Big( {\tau^\prime \over a}
       \Big)^\prime \xi^0,
   \label{sigma-GT}
\eea
and $\Upsilon$, denoted as $\alpha_\sigma$, is gauge-invariant, thus ${\cal J}$ is also gauge-invariant. For $\tau$-fluid quantities, we can show
\bea
   \widehat \mu = \mu - \mu^\prime \xi^0, \quad
       \widehat p = p - p^\prime \xi^0, \quad
       \widehat v = v - c \xi^0,
   \label{mu-GT}
\eea
which are also valid for ordinary fluids. For gauge transformation properties of the metric, see Eq.\ (250) of \cite{Noh-Hwang-2004}.

From Eq.\ (\ref{sigma-GT}) we notice that $\sigma$ can be used to fix the gauge condition, and $\sigma = 0$ in all coordinates gives $\xi^0 = 0$, thus fixes the gauge (coordinate) degree of freedom completely; this implies that under such a gauge condition the remaining perturbation variables are gauge-invariant. Equation (\ref{fluid-linear}) shows that $\sigma = 0$ implies $v = 0$, thus corresponds to the comoving gauge of the $\tau$-fluid. This analysis can be extended to fully nonlinear order, see Sec.\ VI of \cite{Noh-Hwang-2004}.

As a temporal gauge condition we can use $\sigma = 0$ in all coordinate systems, and we may as well call it uniform-$\tau$ gauge. This implies $v = 0$  for the $\tau$-fluid, and we may as well call it the comoving gauge of the $\tau$-fluid. Similarly we can impose $v = 0$ for an ordinary fluid. As another choice we can set $\delta \mu = 0$ in all coordinates which is the uniform-density gauge for the $\tau$-fluid or ordinary fluid. Other temporal gauge conditions which remove the gauge degree of freedom completely are the following: $\varphi \equiv 0$ (uniform-curvature gauge), $\kappa \equiv 0$ (uniform expansion gauge), $\chi \equiv 0$ (zero-shear gauge) \cite{Bardeen-1988}. One exception where the temporal gauge degree of freedom is not completely fixed even after imposing the gauge condition is $\alpha \equiv 0$ (synchronous gauge). These gauge conditions are applicable to nonlinear and exact order with corresponding gauge invariance for other than synchronous gauge \cite{Noh-Hwang-2004, Hwang-Noh-2013}.

\subsubsection{Density perturbation in $\sigma = 0$ gauge}
                                        \label{sec:linear-density}

We consider density perturbation equation of the $\tau$-field in the $\sigma = 0$ gauge condition. The $\sigma = 0$ gauge condition implies the comoving gauge ($v = 0$) of the $\tau$-fluid, which is true even to fully nonlinear order in perturbation, see Eq.\ (\ref{fluid-FNLE}). Under this gauge Eqs.\ (\ref{mom-conserv-linear}) and (\ref{E-conserv-linear}), respectively, give
\bea
   \alpha_\sigma = - {\delta p_\sigma \over \mu + p}, \quad
       \kappa_\sigma = { (a^3 \delta \mu_\sigma
       )^{\displaystyle{\cdot}}
       \over a^3 ( \mu + p )},
   \label{conservation-eqs-tau}
\eea
where we decomposed $\mu = \bar \mu + \delta \mu$ and $p = \bar p + \delta p$. A subindex $\sigma$ indicates the $\sigma = 0$ gauge imposed, or equivalently a unique gauge-invariant combination; from Eqs.\ (\ref{sigma-GT}) and (\ref{mu-GT}), we have
\bea
   \delta \mu_\sigma \equiv \delta \mu
       - {\mu^\prime \over \tau^\prime} {\sigma \over c^2},
\eea
which is a unique gauge-invariant combination of $\delta \mu$ and $\sigma$ and becomes $\delta \mu$ in the $\sigma = 0$ gauge; as long as a gauge condition completely fixes the gauge degree of freedom, all variables in that gauge have unique corresponding gauge-invariant combinations. Ignoring other fluids, Eq.\ (\ref{eq4-linear}) gives
\bea
   {1 + w \over a^2 H} \Big[ {H^2 \over a (\mu + p)}
       \Big( {a^3 \mu \over H} \delta_\sigma
       \Big)^{\displaystyle{\cdot}}
       \Big]^{\displaystyle{\cdot}}
       = c^2 {\Delta \over a^2} {\delta p_\sigma \over \mu},
   \label{delta-eq-linear}
\eea
where $\delta \equiv \delta \mu/\mu$; it is valid in the presence of the cosmological constant. The same equation is valid for a general fluid in the absence of anisotropic stress \cite{HN-GRG-1999}.

The ${\cal K} (Q)$-term is introduced in the Lagrangian in order to reproduce cosmology which often demands presence of dark matter in the context of CMB and the large-scale structure. Here, we are curious whether the pressure terms in the above equation can be ignored in the perturbation level so that the $\tau$-field in the form of ${\cal K}$-term can supplement the dark matter in the large-scale demanded in standard cosmology.

If we can ignore background pressure compared with background energy density, as in the model-I considered in Eq.\ (\ref{constraint-BG}), Eq.\ (\ref{delta-eq-linear}) gives
\bea
   & & \hskip -.6cm
       {1 \over a^2 H} \Big[ a^2 H^2
       \Big( {\delta_\sigma \over H}
       \Big)^{\displaystyle{\cdot}}
       \Big]^{\displaystyle{\cdot}}
       = \ddot \delta_\sigma + 2 H \dot \delta_\sigma
       - 4 \pi G \varrho \delta_\sigma
       = c^2 {\Delta \over a^2} {\delta p_\sigma \over \mu}.
   \nonumber \\
   \label{delta-eq-linear-p=0}
\eea
Comparison of the pressure term with gravity term gives the Jeans scale. Setting $\Delta = - k^2$, we have the Jeans wavenumber
\bea
   {k_J \over a}
       = \sqrt{4 \pi G \delta \varrho_\sigma
       \over c^2 \delta p_\sigma/\mu}
       = \Big[ 2 c_\tau^2 \Big( {w_\tau \over {\cal K}}
       + {a^3 \over I_0} {1 \over Q} {\cal J}_{,A} \Big)
       \Big]^{-1/2},
   \label{Jeans-general}
\eea
where the right-hand side is evaluated using the background parameters; we used Eqs.\ (\ref{EOM-BG}), (\ref{w-cs}), (\ref{Q-linear}), (\ref{Upsilon}) and (\ref{fluid-linear}). The pressure perturbation can be ignored in super-Jeans scale, and the density perturbation of $\tau$-field behaves as a cold dark matter for negligible pressure perturbation with $k \ll k_J$. The above Jeans criterion is valid for general ${\cal J}$ and ${\cal K}$, assuming the background $\tau$-field behaves as a dust fluid.

\subsubsection{Perturbation constraint}
                                        \label{sec:constraint-pert}

For the model-I in Eq.\ (\ref{model-I}), we have the background equation of state in Eq.\ (\ref{BG-sols}). For perturbed part, using Eq.\ (\ref{fluid-linear}), we have
\bea
   & & \delta \mu_\sigma = - {c^4 \over 8 \pi G}
       \Big( n {I_0 \over a^3} {Q^2 \over Q -1}
       + 2 {\cal J}_{,A} {\Delta \over a^2} \Big) \alpha,
   \nonumber \\
   & & \delta p_\sigma = - {c^4 \over 8 \pi G} {I_0 \over a^3}
       Q \alpha.
\eea
Equation (\ref{Jeans-general}) gives
\bea
   {k_J \over a}
       = \Big[ {2 a^3 \over I_0}
       {Q - 1 \over n Q^2}
       \Big( {(n + 1) Q \over n Q + 1} + {\cal J}_{,A} \Big) \Big]^{-1/2}.
   \label{Jeans-I-general}
\eea
For $Q \simeq 1$, using Eq.\ (\ref{BG-sols}), we have
\bea
   & & {\delta p_\sigma \over \mu}
       = c_\tau^2 \Big( 1 - {\cal J}_{,A}
       {c_\tau^2 c^2 k^2 \over 4 \pi G \varrho_\tau a^2}
       \Big)^{-1} \delta_\sigma,
   \nonumber \\
   & & {c_\tau^2 c^2 k_J^2 \over 4 \pi G \varrho_\tau a^2}
       ( 1 + {\cal J}_{,A} ) = 1.
\eea
The Jeans scale can be written as
\bea
   {k_J \over a}
       \simeq \Big[ {2 \over n} {a^3 \over I_0}
       \Big( {I_0 \over 2 \nu^2 {\cal K}_{n+1} a^3}
       \Big)^{1/n}
       (1 + {\cal J}_{,A} ) \Big]^{-1/2}.
   \label{Jeans-I}
\eea

For $n = 1$, we have
\bea
   {k_J \over a}
       \simeq {\nu \over \sqrt{ 1 + {\cal J}_{,A} } },
\eea
thus, the Jeans scale corresponds to the length scale of the mass-like term. In Eq.\ (\ref{constraint-MOND}) a reasonable phenomenology in galactic scale demanded $1/\nu > 1{\rm Mpc}$ and this applies only for $n = 1$. In Eq.\ (\ref{constraint-BG}) a dust-like behavior in the background demands $1/\nu \le 220 pc$. Thus, $n = 1$ case is excluded \cite{Blanchet-Skordis-2024}.

For $n \geq 2$, we have no constraint in Eq.\ (\ref{constraint-MOND}). For example, for $n = 2$, we have
\bea
   {k_J \over a}
       \simeq \sqrt{ \sqrt{2 I_0 \over a^3}
       {\nu \sqrt{ {\cal K}_3 } \over 1 + {\cal J}_{,A}} }.
\eea
We can estimate
\bea
   & & {\lambda_J \over 2 \pi} = {a \over k_J}
       \leq \sqrt{ {L_H L_* a^{3/2}
       \over \sqrt{6 \Omega_{\tau0}}} ( 1 + {\cal J}_{,A} ) }
   \nonumber \\
   & & \qquad
       \sim \sqrt{1 + {\cal J}_{,A}} a^{3/4} 0.46 Mpc,
   \label{constraint-pert}
\eea
where we used $L_H \equiv c/H_0 = 4.3 Gpc$ and Eq.\ (\ref{constraint-BG}). In Newtonian regime we may set ${\cal J}_{,A} = 0$ whereas in the MOND regime we have $1 + {\cal J}_{,A} = {1 \over a} |\nabla \Phi|/a_M$. Above the Jeans scale the density perturbation of $\tau$-field behaves as a cold DM.

\begin{widetext}
\subsection{Fully nonlinear and exact perturbations in $\sigma = 0$ gauge}
                                        \label{sec:FNLE}

Setting $\sigma \equiv 0$ as the temporal gauge condition, Eq.\ (\ref{Q-perturbation}) significantly simplifies to become
\bea
   Q = {1 \over {\cal N}} \dot \tau^2, \quad
       {\cal N} \equiv \sqrt{ 1 + 2 \alpha
       + {\chi^k \chi_k \over a^2 (1 + 2 \varphi)}},
   \label{Q-FNLE}
\eea
where $\tau = \bar \tau (t)$. The four-vector becomes
\bea
   U_0 = - a {\cal N}, \quad
       U_i = 0, \quad
       U^0 = {1 \over a {\cal N}}, \quad
       U^i = {\chi^i \over a^2 {\cal N} (1 + 2 \varphi)}.
\eea
This is the same as the normal four-vector in Eq.\ (93) of \cite{Hwang-Noh-EM-2023}. The acceleration vector becomes
\bea
   & & A_0 = {c^2 \chi^i \over a (1 + 2 \varphi)}
       {Q_{,i} \over Q}, \quad
       A_i = - c^2 {Q_{,i} \over Q}, \quad
       A^0 = 0, \quad
       A^i = - {c^2 \over a^2 (1 + 2 \varphi)}
       {Q^{,i} \over Q}, \quad
       A = {Q^{,i} Q_{,i} \over a^2 (1 + 2 \varphi) Q^2}.
\eea
The energy-momentum tensor of the $\tau$-fluid becomes
\bea
   & & T_{00} = {c^4 \over 8 \pi G} \Big[ a^2 (1 + 2 \alpha)
       ( {\cal J} - {\cal K} )
       + 2 {\cal J}_{,A} \Big( { \chi^i Q_{,i}
       \over a ( 1 + 2 \varphi ) Q} \Big)^2
       + {2 {\cal N} \over (1 + 2 \varphi)^{3/2}}
       \Big( {\cal N} \sqrt{1 + 2 \varphi} {\cal J}_{,A} {Q^{,i}
       \over Q} \Big)_{,i}
       + a^2 {\cal N}^2 Q {\cal K}_{,Q} \Big],
   \nonumber \\
   & & T_{0i} = {c^4 \over 8 \pi G} \Big[
       ( {\cal J} - {\cal K} ) a \chi_i
       - 2 {\cal J}_{,A} {\chi^j Q_{,j} Q_{,i}
       \over a ( 1 + 2 \varphi ) Q^2} \Big], \quad
       T_{ij} = {c^4 \over 8 \pi G} \Big[
       - a^2 ( 1 + 2 \varphi ) ( {\cal J} - {\cal K} )
       \delta_{ij}
       + 2 {\cal J}_{,A} {Q_{,i} Q_{,j} \over Q^2} \Big].
\eea
Using Eqs.\ (\ref{fluid-ADM-1}) and (\ref{fluid-ADM-2}), the $\tau$-fluid quantities become
\bea
   & & E = {c^4 \over 8 \pi G}
       \Big[ {\cal J} - {\cal K} + Q {\cal K}_{,Q}
       + {2 \over {\cal N} (1 + 2 \varphi)^{3/2}}
       \Big( {\cal N} \sqrt{1 + 2 \varphi} {\cal J}_{,A} {Q^{,i}
       \over Q} \Big)_{,i} \Big], \quad
       m_i = 0,
   \nonumber \\
   & & m_{ij} = {c^4 \over 8 \pi G} \Big[
       - ( {\cal J} - {\cal K} ) \delta_{ij}
       + {2 {\cal J}_{,A} \over a^2 (1 + 2 \varphi)}
       {Q_{,i} Q_{,j} \over Q^2} \Big], \quad
       \overline m_{ij} = {c^4 \over 8 \pi G}
       {2 {\cal J}_{,A} \over a^2 ( 1 + 2 \varphi ) Q^2}
       \Big( Q_{,i} Q_{,j} - {1 \over 3} Q^{,k} Q_{,k} \delta_{ij} \Big).
   \label{fluid-FNLE}
\eea
$m_i = 0$ shows that for $\sigma = 0$ gauge condition the velocity of the $\tau$-fluid vanishes, thus comoving to fully nonlinear order.

Einstein's equation and the conservation equations in the fully nonlinear and exact formulation with the above fluid quantities are presented in Eqs.\ (C8)-(C12), (C14) and (C15) of \cite{Hwang-Noh-EM-2023}. Einstein's equations give
\bea
   & & - {3 \over 2} \Big( {\dot a^2 \over a^2}
       - {8 \pi G \over 3 c^2} E
       - {\Lambda c^2 \over 3} \Big)
       + {\dot a \over a} \kappa
       + {c^2 \Delta \varphi \over a^2 (1 + 2 \varphi)^2}
       = {1 \over 6} \kappa^2
       + {3 \over 2} {c^2 \varphi^{,i} \varphi_{,i} \over a^2 (1 + 2 \varphi)^3}
       - {c^2 \over 4} \overline{K}^i_j \overline{K}^j_i,
   \label{eq2-FNLE} \\
   & & {2 \over 3} {1 \over c} \kappa_{,i}
       + {1 \over a^2 {\cal N} ( 1 + 2 \varphi )}
       \Big( {1 \over 2} \Delta \chi_i
       + {1 \over 6} \chi^k_{\;\;,ki} \Big)
       =
       {1 \over a^2 {\cal N} ( 1 + 2 \varphi)}
       \Big\{
       \Big( {{\cal N}_{,j} \over {\cal N}}
       - {\varphi_{,j} \over 1 + 2 \varphi} \Big)
       \Big[ {1 \over 2} \Big( \chi^{j}_{\;\;,i} + \chi_i^{\;,j} \Big)
       - {1 \over 3} \delta^j_i \chi^k_{\;\;,k} \Big]
   \nonumber \\
   & & \qquad
       - {\varphi^{,j} \over (1 + 2 \varphi)^2}
       \Big( \chi_{i} \varphi_{,j}
       + {1 \over 3} \chi_{j} \varphi_{,i} \Big)
       + {{\cal N} \over 1 + 2 \varphi} \nabla_j
       \Big[ {1 \over {\cal N}} \Big(
       \chi^{j} \varphi_{,i}
       + \chi_{i} \varphi^{,j}
       - {2 \over 3} \delta^j_i \chi^{k} \varphi_{,k} \Big) \Big]
       \Big\}
       - {8 \pi G \over c^3} a m_i,
   \label{eq3-FNLE} \\
   & & - 3 \Big[ {1 \over {\cal N}}
       \Big( {\dot a \over a} \Big)^{\displaystyle\cdot}
       + {\dot a^2 \over a^2}
        + {4 \pi G \over 3 c^2} \Big( E + S \Big)
       - {\Lambda c^2 \over 3} \Big]
       + {1 \over {\cal N}} \dot \kappa
       + 2 {\dot a \over a} \kappa
       + {c^2 \Delta {\cal N} \over a^2 {\cal N} (1 + 2 \varphi)}
   \nonumber \\
   & & \qquad
       = {1 \over 3} \kappa^2
       - {c \over a^2 {\cal N} (1 + 2 \varphi)} \Big(
       \chi^{i} \kappa_{,i}
       + c {\varphi^{,i} {\cal N}_{,i} \over 1 + 2 \varphi} \Big)
       + c^2 \overline{K}^i_j \overline{K}^j_i,
   \label{eq4-FNLE} \\
   & & \Big( {1 \over {\cal N}} {\partial \over \partial t}
       + 3 {\dot a \over a}
       - \kappa
       + {c \chi^{k} \over a^2 {\cal N} (1 + 2 \varphi)} \nabla_k \Big)
       \Big\{ {c \over a^2 {\cal N} (1 + 2 \varphi)}
   \nonumber \\
   & & \qquad
       \times
       \Big[
       {1 \over 2} \Big( \chi^i_{\;\;,j} + \chi_j^{\;\;,i} \Big)
       - {1 \over 3} \delta^i_j \chi^k_{\;\;,k}
       - {1 \over 1 + 2 \varphi} \Big( \chi^{i} \varphi_{,j}
       + \chi_{j} \varphi^{,i}
       - {2 \over 3} \delta^i_j \chi^{k} \varphi_{,k} \Big)
       \Big] \Big\}
   \nonumber \\
   & & \qquad
       - {c^2 \over a^2 ( 1 + 2 \varphi)}
       \Big[ {1 \over 1 + 2 \varphi}
       \Big( \nabla^i \nabla_j - {1 \over 3} \delta^i_j \Delta \Big) \varphi
       + {1 \over {\cal N}}
       \Big( \nabla^i \nabla_j - {1 \over 3} \delta^i_j \Delta \Big) {\cal N} \Big]
   \nonumber \\
   & & \qquad
       =
       - {c^2 \over a^2 (1 + 2 \varphi)^2}
       \Big[ {3 \over 1 + 2 \varphi}
       \Big( \varphi^{,i} \varphi_{,j}
       - {1 \over 3} \delta^i_j \varphi^{,k} \varphi_{,k} \Big)
       + {1 \over {\cal N}} \Big(
       \varphi^{,i} {\cal N}_{,j}
       + \varphi_{,j} {\cal N}^{,i}
       - {2 \over 3} \delta^i_j \varphi^{,k} {\cal N}_{,k} \Big) \Big]
   \nonumber \\
   & & \qquad
       + {c^2 \over a^4 {\cal N}^2 (1 + 2 \varphi)^2}
       \Big[
       {1 \over 2} \Big( \chi^{i,k} \chi_{j,k}
       - \chi_{k,j} \chi^{k,i} \Big)
       + {1 \over 1 + 2 \varphi} \Big(
       \chi^{k,i} \chi_k \varphi_{,j}
       - \chi^{i,k} \chi_j \varphi_{,k}
       + \chi_{k,j} \chi^k \varphi^{,i}
       - \chi_{j,k} \chi^i \varphi^{,k} \Big)
   \nonumber \\
   & & \qquad
       + {2 \over (1 + 2 \varphi)^2} \Big(
       \chi^{i} \chi_{j} \varphi^{,k} \varphi_{,k}
       - \chi^{k} \chi_{k} \varphi^{,i} \varphi_{,j} \Big) \Big]
       + {8 \pi G \over c^2} \overline m^i_j,
   \label{eq5-FNLE}
\eea
where
\bea
   & & \kappa
       \equiv
       3 {\dot a \over a} \Big( 1 - {1 \over {\cal N}} \Big)
       - {1 \over {\cal N} (1 + 2 \varphi)}
       \Big[ 3 \dot \varphi
       + {c \over a^2} \Big( \chi^k_{\;\;,k}
       + {\chi^{k} \varphi_{,k} \over 1 + 2 \varphi} \Big)
       \Big],
   \label{eq1-FNLE} \\
   & & \overline{K}^i_j \overline{K}^j_i
       = {1 \over a^4 {\cal N}^2 (1 + 2 \varphi)^2}
       \Big\{
       {1 \over 2} \chi^{i,j} \Big( \chi_{i,j} + \chi_{j,i} \Big)
       - {1 \over 3} \chi^i_{\;\;,i} \chi^j_{\;\;,j}
       - {4 \over 1 + 2 \varphi} \Big[
       {1 \over 2} \chi^i \varphi^{,j} \Big(
       \chi_{i,j} + \chi_{j,i} \Big)
       - {1 \over 3} \chi^i_{\;\;,i} \chi^j \varphi_{,j} \Big]
   \nonumber \\
   & & \qquad
       + {2 \over (1 + 2 \varphi)^2} \Big(
       \chi^{i} \chi_{i} \varphi^{,j} \varphi_{,j}
       + {1 \over 3} \chi^i \chi^j \varphi_{,i} \varphi_{,j} \Big) \Big\}.
   \label{K-bar-eq}
\eea
To the linear order Eqs.\ (\ref{eq2-FNLE})-(\ref{eq1-FNLE}) gives
Eqs.\ (\ref{eq2-linear})-(\ref{eq5-linear}) and (\ref{eq1-linear}), respectively. For fully nonlinear and exact formulation including the transverse (vector) and transverse-tracefree (tensor) modes, see \cite{Gong-2017}. The fluid quantities in Einstein's equations are the collective ones, for example, $E = E_m + E_\tau$, $m_i = m_i^m + m_i^\tau$, $m_{ij} = m_{ij}^m + m_{ij}^\tau$, etc; for the $\tau$-fluid, fluid quantities are given in Eq.\ (\ref{fluid-FNLE}), and for ordinary fluid, scalar field, and Maxwell field, see Sec.\ IV.C in \cite{Hwang-Noh-EM-2023}.

The energy and momentum conservation equations give
\bea
   & & {1 \over a^3} \Big[
       a^3 \Big( 1 + 2 \varphi \Big)^{3/2} E
       \Big]^{\displaystyle\cdot}
       + {1 \over a} \Big[
       \Big( 1 + 2 \varphi \Big)^{1/2}
       \Big( {\cal N} c^2 m^i
       + {c \over a} \chi^i E \Big) \Big]_{,i}
       = - {1 \over a} \Big( 1 + 2 \varphi \Big)^{1/2}
       {\cal N}_{,i} c^2 m^i
   \nonumber \\
   & & \qquad
       - \Big( 1 + 2 \varphi \Big)^{1/2}
       \Big[ \Big( {\dot a \over a}
       + \dot \varphi + 2 {\dot a \over a} \varphi \Big)
       \delta^{ij}
       + {c \over a^2} \chi^{i,j}
       - {c \over a^2 (1 + 2 \varphi)}
       \Big( 2 \chi^i \varphi^{,j}
       - \delta^{ij} \chi^k \varphi_{,k} \Big) \Big]
       m_{ij},
   \label{eq6-FNLE} \\
   & & {1 \over a^4} \Big[ a^4
       \Big( 1 + 2 \varphi \Big)^{3/2} m_i
       \Big]^{\displaystyle\cdot}
       + {1 \over a} \Big[
       \Big( 1 + 2 \varphi \Big)^{3/2}
       \Big( {\cal N} m_i^j
       + {c \chi^j m_i \over a (1 + 2 \varphi)} \Big)
       \Big]_{,j}
   \nonumber \\
   & & \qquad
       = {1 \over a} \Big( 1 + 2 \varphi \Big)^{3/2} \Big[
       {{\cal N} \varphi_{,i} \over 1 + 2 \varphi} S
       - {\cal N}_{,i} E
       - {c \over a} \Big( {\chi^j \over 1 + 2 \varphi} \Big)_{,i} m_j \Big].
   \label{eq7-FNLE}
\eea
The conservation equations are valid for $\tau$-fluid and ordinary fluids, including scalar field as well as Maxwell field \cite{Hwang-Noh-EM-2023}. The equation of motion in Eq.\ (\ref{EOM}) gives
\bea
   & & {1 \over c} {\partial \over \partial t}
       \Big[ {a \over {\cal N} Q}
       \Big( {\cal N} \sqrt{1 + 2 \varphi}
       {\cal J}_{,A} {Q^{,i} \over Q} \Big)_{,i}
       + {1 \over 2} a^3 ( 1 + 2 \varphi )^{3/2}
       {\cal K}_{,Q} \Big]
       + {\partial \over \partial x^i} \Big\{
       \sqrt{1 + 2 \varphi}
       {\cal J}_{,A} {Q^{,i} \over Q^3}
       \Big( {a \over c} \dot Q
       + {\chi^j Q_{,j} \over a (1 + 2 \varphi)} \Big)
   \nonumber \\
   & & \qquad
       + {\chi^i \over a (1 + 2 \varphi)}
       \Big[ {1 \over {\cal N} Q} \Big( {\cal N}
       \sqrt{1 + 2 \varphi}
       {\cal J}_{,A} {Q^{,j} \over Q} \Big)_{,j}
       + {1 \over 2} a^2 ( 1 + 2 \varphi )^{3/2}
       {\cal K}_{,Q} \Big] \Big\}
       = 0.
   \label{EOM-FNLE}
\eea
\end{widetext}
Using Eq.\ (\ref{fluid-FNLE}), we can show that, for $\tau$-fluid,  Eq.\ (\ref{eq6-FNLE}) is the same as Eq.\ (\ref{EOM-FNLE}) and Eq.\ (\ref{eq7-FNLE}) is identically valid.

Equations (\ref{eq2-FNLE})-(\ref{EOM-FNLE}) are the complete set of {\it exact} perturbation equations. The conservation equations in (\ref{eq6-FNLE})-(\ref{eq7-FNLE}) are valid for fluids as well as for $\tau$-field with the equation of state presented in Eq.\ (\ref{fluid-FNLE}); for the field we can also use the equation of motion in Eq.\ (\ref{EOM-FNLE}).

%
%
%
\section{Discussion}
                                        \label{sec:discussion}

A simple relativistic extension of MOND theory and its minimal modification to account for cosmological demands for DM was recently proposed by Blanchet, Marsat and Skordis \cite{Blanchet-Marsat-2011, Blanchet-Skordis-2024}; we call it the BMS theory. The theory is summarized in Eqs.\ (\ref{L}) and (\ref{Q}). Here, we derived equations of the PN approximation and perturbation theory in cosmological context.

The PN approximation of the BMS theory was presented in \cite{Flanagan-2023, Blanchet-Skordis-2024} in Minkowsky background, and the cosmological linear perturbation was presented in \cite{Blanchet-Skordis-2024}. Compared with previous works, here we presented equations valid to 1PN order in cosmology context, linear perturbations without imposing temporal gauge condition, and the fully-nonlinear and exact perturbation equations. We clarified the gauge issues in both formulations. We presented the 1PN approximation assuming $\tau = t$ as a physical condition and fully nonlinear and exact perturbations assuming $\tau = \tau (t)$ using a temporal gauge condition. General cases without imposing these conditions demand lengthy presentations, but the methods are the same.

In linear perturbations, we kept general ${\cal J}$ and ${\cal K}$ and derived Jeans criterion for a constraint on $\tau$-fluid. The Bekenstein and Milgrom formulation of MOND \cite{Bekenstein-Milgrom-1984} can be recovered by using ${\cal J} (A)$ in the 0PN limit of the PN approximation by taking $\sigma = 0$ as a physical condition on $\tau$-field, thus $\tau = t$. 

Constraints on ${\cal K} (Q)$ follow from the background, perturbations, and the MOND requirements. (i) A demand for dust-like equation of state for $\tau$-field in the background constrains ${\cal K}$: see Eq.\ (\ref{constraint-BG}) for model-I. (ii) For model-II, a mass-like term appears for $n = 1$, and the upper limit on this term gives a constraint in Eq.\ (\ref{constraint-MOND}) which conflicts with the background constraint. These two constraints were presented in \cite{Blanchet-Skordis-2024}. (iii) A demand for $\tau$-fluid to behave as a pressureless fluid in linear density perturbation leads to another constraint. The Jeans scale is derived in Eq.\ (\ref{Jeans-general}) for general ${\cal J}$ and ${\cal K}$, and in Eq.\ (\ref{Jeans-I}) for model-I. Proper estimation of the effect of ${\cal K}$ in cosmological linear perturbations needs numerical study of the full system summarized below Sec.\ \ref{sec:linear}.

In BMS theory, the two functions ${\cal J}$ and ${\cal K}$ are still free to accommodate the phenomena in MOND and successful cosmology. Future studies may design and constrain the functional forms using theories and observations. Other functional forms, especially ${\cal K}$ for cosmology were previously considered with successful fits to the power spectra of the large-scale matter and CMB \cite{Blanchet-Skordis-2024, Skordis-Zlosnik-2021}. The fully nonlinear and exact perturbation equations will be useful to study the leading nonlinear matter power spectrum  as well as the non-Gaussian effects.

Recent JWST observation of the early emergence of massive galaxies in high redshifts is regarded as a challenge to the cold dark matter (CDM) cosmology \cite{Carniani-2024}. The early formation of galaxies was a predicted feature in the MOND model \cite{Sanders-1998, McGaugh-2024}. In the low-acceleration MOND regime the gravity is stronger than Newtonian gravity, thus the structures can grow faster \cite{Nusser-2002}. We showed the case in the BMS model using the 0PN approximation in Section \ref{sec:MOND-growth}.

We point out a couple of caveats used in our PN approximation. Our PN approximation is limited by adopting $\tau = t$ to the background order, thus excluding proper consideration of the $\tau$-field contribution to the Friedmann background. Adopting $\sigma = 0$ as a physical condition to 0PN order is another, a sort of, ansatz limiting the potential role of the $\tau$-field. The first condition limits our study valid in the Friedmann background without ${\cal K}$ contribution, see below Eq.\ (\ref{tau-PN}). The latter condition conveniently leads us to the MOND phenomenology but its meaning should be clarified. These are interesting issues demanding further investigation by relaxing the conditions.

We showed that above Jeans scale the density perturbation of $\tau$-field behaves as a pressureless fluid. To confirm whether the Jeans scale in Eq.\ (\ref{Jeans-general}) [Eq.\ (\ref{Jeans-I}) for the model-I and  Eq.\ (\ref{constraint-pert}) for $n = 2$] is sufficient to reproduce the large-scale structure and CMB observations which are consistent with the $\Lambda$CDM paradigm, we may need numerical study by including contributions of the baryon, photons and other cosmological paraphernalia.

If successful, by adding two functions of a time-like vector field in a gradient form, the BMS theory encompasses the MOND phenomenon {\it within} Einstein's gravity paradigm algebraically, although the local Lorentz invariance and the strong equivalence principle are broken conceptually. The latter violations may open new possibilities to test the theory.

\vskip .2cm
%
%
%
\centerline{\bf Acknowledgments}

We thank an anonymous referee for useful suggestions. H.N.\ was supported by the National Research Foundation (NRF) of Korea funded by the Korean Government (No.RS-2024-00333721 and No.2021R1F1A1045515). J.H.\ was supported by IBS under the project code, IBS-R018-D1.

%
%


\end{document}